\documentclass[superscriptaddress, twocolumn, a4paper, tightenlines, nofootinbib, preprintnumbers]{revtex4-2}

\usepackage{xcolor}
\usepackage{graphicx}
\usepackage{bm}
\usepackage{amssymb,amsmath,amsfonts,bbm}
\usepackage{caption}
\usepackage{subcaption}
\usepackage[normalem]{ulem}

\usepackage{hyperref}
\hypersetup{colorlinks=true,allcolors=red}

\newcommand{\bV}{\mathbf{V}}
\newcommand{\bX}{\mathbf{X}}

\begin{document}

\title{Hamiltonian curl-forces in systems coupled to multiple thermal reservoirs}

\author{Omer Chor}
\affiliation{Department of Physics of Complex Systems, Weizmann Institute of Science, Rehovot, 761001, Israel}
\author{Alja\v{z} Godec}
\affiliation{Mathematical Physics and Stochastic Dynamics, Institute of Physics,
University of Freiburg, 79104 Freiburg im Breisgau, Germany}
\affiliation{Mathematical bioPhysics Group, Max Planck Institute for Multidisciplinary Sciences, G\"ottingen 37077, Germany}
\author{Oren Raz}
\affiliation{Department of Physics of Complex Systems, Weizmann Institute of Science, Rehovot, 761001, Israel}

\date{\today}

\begin{abstract}
We investigate the statistical mechanics of a system coupled to multiple heat reservoirs at different temperatures, a setting which typically sustains nontrivial circulation. By applying a canonical transformation, we map the system onto that of a particle coupled to a single reservoir yet driven by a deterministic curl-force---a reversible, velocity-independent non-conservative force with non-vanishing curl---that can be described by a Hamiltonian with an anisotropic kinetic energy term. Despite the Hamiltonian structure, the system does not relax to the corresponding Boltzmann distribution, and is genuinely out of equilibrium. The mapping provides a mechanical interpretation of the currents in multi-temperature systems, as we demonstrate by means of two examples. In particular, the direction of circulation is qualitatively predictable from the deterministic force field. For the exactly solvable case of an $N$-dimensional underdamped Ornstein-Uhlenbeck process, we derive explicit expressions for the steady-state distribution and the entropy production rate, highlighting the role of the (generalized) curl of the force field in driving the system out of thermal equilibrium. Finally, we 
show how a non-conservative force can be used to exactly equilibrate any system coupled to multiple reservoirs, and obtain the corresponding equilibrium distribution.
\end{abstract}

\maketitle
The study of systems coupled to multiple reservoirs dates back to the foundations of classical thermodynamics, where the heat engine operating between a hot and cold reservoir, as analyzed by Carnot, serves as the paradigmatic setup \cite{carnot1872reflexions}. Clausius's formulation of the second law of thermodynamics also concerns heat flows between reservoirs \cite{clausius1879mechanical}. 
Systems coupled to multiple baths remain of great interest to date, where typically a microscopic model is considered instead of the earlier macroscopic descriptions. These include boundary-driven models such as asymmetric simple exclusion processes (ASEP) \cite{krug1991boundary, derrida1998exactly, derrida2007non} or oscillator chains \cite{kipnis1982heat, lepri1997heat, eckmann1999non}. More recently, there has also been great interest in systems where different degrees of freedom (DoF) are coupled to a different heat bath within the framework of stochastic thermodynamics \cite{esposito2012stochastic, seifert2012stochastic, murashita2016overdamped}. Notable examples are Brownian motors, refrigerators and gyrators \cite{van2004microscopic, van2006brownian, filliger2007brownian}, which are a modern take on the classical macroscopic heat engine. These ideas have been generalized and studied extensively both theoretically and experimentally \cite{ dotsenko2013two, mancois2018two, bae2021inertial, baldassarri2020engineered,cerasoli2018asymmetry, cerasoli2022spectral, chang2021autonomous, dos2021stationary, pietzonka2018universal, movilla2021energy, abdoli2022tunable, chiang2017electrical, ciliberto2013heat, PhysRevE.92.032118, abdoli2025quadrupolar, visco2006work, fogedby2011bound, fogedby2017minimal, leighton2024information, raz2016geometric, netz2020approach, lin2022stochastic, argun2017experimental, martins2026role, berut2014energy, berut2016stationary, wu2022nonequilibrium} in the past two decades. In these systems, the simultaneous coupling to several reservoirs imposes incompatible equilibrium conditions, breaking detailed balance. Some nonequilibrium systems may never reach a steady state at all. But even the seemingly simpler case of a nonequilibrium steady state (NESS) poses a significant theoretical challenge  \cite{seifert2012stochastic, sasa2006steady, zia2006possible, zia2007probability, wang2008potential}, because there is no direct analog of the universal equilibrium (Boltzmann-Gibbs) 
distribution for a NESS. This means that exact solutions typically require a case-by-case analytical approach.

In addition to coupling to multiple reservoirs, there exist other ways of driving a system out of equilibrium. In particular, we will be interested in systems influenced by some external non-conservative force \cite{seifert2012stochastic, sasa2006steady, chernyak2006path,busiello2018similarities}. In principle, all the fundamental forces in nature are conservative, and can be derived from a potential energy function. However, in cases where a complete microscopic description is unavailable for some parts of the total system, or in cases where a faster ``hidden" process coupled to the system (e.g.\ a chemical reaction) locally injects energy into the system, it  is often useful to  effectively describe the dynamics through non-conservative forces. For example, drag forces are often introduced to account for the influence of a medium on the motion of an immersed particle. Conversely, non-potential forces are included to account for the transduction of chemical energy into mechanical work \cite{hill2012free} or as a result of hydrodynamic flows advecting a particle \cite{Foister_Ven_1980}.
Generally, non-conservative forces can be \emph{dissipative}, which means that the dynamics of the system are not reversible and do not conserve phase space volume. Such non-conservative forces do not necessarily violate the detailed balance condition (and evidently, they are necessary for the equilibration of an excited system). However, there also exist reversible non-conservative forces, which conserve phase space volume but are not derived from a potential. Specifically, we are interested in the case of force fields $\mathbf{F}(\mathbf{x})$ that depend only on the position $\mathbf{x}$ of the particles but not on their velocities, and are not the gradient of any scalar potential. In three dimensions, this implies that they have a non-vanishing curl, $\nabla \times \mathbf{F}(\mathbf{x})\neq0$. Following Berry and Shukla, these are referred to as \emph{curl-forces} \cite{berry2012classical}. 

{In this work we uncover 
a mapping between  
two mechanisms of driving a system out of thermal equilibrium: applying a non-conservative curl-force, and connecting it to multiple heat reservoirs. We show how  this mapping not only allows for a formal unification of the two types of non-equilibrium settings, but also that the curl-force description of the multi-reservoir system can be used as a practical tool to gain  
understanding and intuition about the system's dynamics.}

The manuscript is organized as follows. We begin by introducing the Langevin and corresponding Klein-Kramers equations describing the underdamped stochastic evolution of systems coupled to multiple heat reservoirs. Then, we present a canonical transformation that maps multiple thermal reservoirs to a single heat bath while fully preserving 
the nonequilibrium nature of the system: 
the transformed system is coupled to a single heat bath but experiences a non-conservative force. Importantly, this force can be described by a non-standard Hamiltonian, and is identified as a \emph{Hamiltonian curl-force} \cite{berry2015hamiltonian}. We include a brief introduction to such forces, discuss the properties of the transformed single-reservoir system, and explain how come its steady state is not an equilibrium state despite the underlying Hamiltonian structure. 
To demonstrate the usefulness of the introduced transformation, we present several applications: in particular 
a harmonic Brownian gyrator \cite{filliger2007brownian}, a gyrator in an azimuthal-symmetric potential \cite{abdoli2025quadrupolar}, and a general $N$-dimensional underdamped Ornstein-Uhlenbeck (OU) process  coupled to $N$ reservoirs \cite{risken1996fokker}. The emerging non-conservative Hamiltonian force allows for an intuitive mechanical interpretation of the circulation and entropy production observed in these systems. We also show how applying a non-conservative force may actually equilibrate a non-equilibrium system 
coupled to multiple reservoirs, and derive the corresponding equilibrium distribution even though, notably, the system has no single inverse temperature so the standard Boltzmann distribution is ill defined. Finally, we summarize our results and provide an outlook with open questions and possible generalizations.

\section*{Underdamped particle coupled to multiple reservoirs}\label{sec:underdamped}
{Consider a classical open system with $2N$
degrees of freedom (DoF) $(\textbf{x},\textbf{p})=(x_1,\ldots x_N; \allowbreak p_1,\ldots,p_N)$
(e.g., an underdamped Brownian 
particle in $N$ dimensions) 
simultaneously connected to multiple heat reservoirs}, such that each momentum DoF 
is coupled to a different reservoir with temperature $T_i$ for $i=1,2,\ldots,N$. In addition, the 
system may 
be subject to a conservative force $-\nabla_\mathbf{x} U(\mathbf{x})$. The particle's energy is given by the Hamiltonian
\begin{equation}
    H(\mathbf{x}, \mathbf{p}) = \frac{\mathbf{p}^2}{2m} + U(\mathbf{x}),
    \label{Hamiltonian_0}
\end{equation}
where $m$ is the system's mass (which, for simplicity and without loss of generality, we assume to be a scalar) and $\mathbf{p}=m\mathbf{v}=m\dot{\mathbf x}$ its momentum. The particle's stochastic equations of motion due to the contact with the bath are described by the Langevin equation \cite{risken1996fokker, sekimoto_stochastic_2010}
\begin{equation}\label{eq:langevin}
    \begin{split}
        d\mathbf{x}(t) &= \mathbf{v}(t)dt \\
        md\mathbf{v}(t) &= -\nabla_{\mathbf{x}} U(\mathbf{x})|_{\mathbf{x} = \mathbf{x}(t)}{dt} - \xi\mathbf{v}(t){dt} + \sqrt{2\xi k_B \mathcal{T}}\cdot d\mathbf{W}(t)
    \end{split}
\end{equation}
where $\mathcal{T} = \mathrm{diag}(T_1, \ldots, T_N)$ is the temperatures matrix, $\xi$ is the friction coefficient (which for simplicity is assumed to be identical for all reservoirs) and $\mathbf{W}(t)$ is the Wiener process
($\boldsymbol{\eta}(t)=``\dot{\mathbf{W}}(t)"$ is a Gaussian white noise) with zero
mean and unit covariance,
\begin{equation}
    \label{eq:white_noise}
    \langle d\mathbf{W}_i(t)\rangle = 0,
    \qquad
    \langle d\mathbf{W}_i(t)d\mathbf{W}_j(t')\rangle = \delta_{ij}\delta(t-t')dtdt'.
\end{equation}
The probability density $f_t(\mathbf{x},\mathbf{v})$ of finding the 
system at a given position $\mathbf{x}$ and velocity $\mathbf{v}$ at time $t$ is obtained by averaging over realizations of the thermal noise. Its evolution---a continuity equation---is governed by the Klein-Kramers
equation~\cite{risken1996fokker}, 
\begin{equation}\label{eq:klein_kramers}
\begin{split}
      \frac{\partial f_t(\mathbf{x}, \mathbf{v})}{\partial t} = &\frac{\nabla_{\mathbf{x}} U\cdot \nabla_{\mathbf{v}}f_t}{m}-\nabla_{\mathbf{x}} f_t \cdot \mathbf{v} \\ &\qquad +  \frac{\xi}{m} \nabla_{\mathbf v} \cdot\left(\mathbf{v} f_t + \frac{k_{\rm B}}{m} \mathcal{T}\cdot\nabla_{\mathbf{v}}f_t\right)\\
  =& \{H,f_t\}+ \frac{\xi}{m} \nabla_{\mathbf v} \cdot\left(\mathbf{v} f_t + \frac{k_{\rm B}}{m} \mathcal{T}\cdot\nabla_{\mathbf{v}}f_t\right).
\end{split}
\end{equation}
Here, $\nabla_{\mathbf{x}}$ and $\nabla_{\mathbf v}$ are the del operators with respect to the coordinates and velocities, respectively. The first two terms, $\frac{\nabla_{\mathbf{x}} U\cdot \nabla_{\mathbf{v}}f_t}{m}-\nabla_{\mathbf{x}} f_t \cdot \mathbf{v}$  describe the (deterministic) Liouville flow of $f_t$, and can be written using the Poisson bracket of the Hamiltonian and the probability density $f_t$, namely as $\{H,f_t\}$. The additional two terms, proportional to $\xi/m$, describe the interaction of the system with the thermal baths (i.e., dissipation and diffusion).

In general, \eqref{eq:klein_kramers} describes a nonequilibrium stochastic process, and exact solutions are typically unknown \cite{risken1996fokker}. The nonequilibrium nature of the process arises from the coupling to multiple baths with different temperatures, such that heat can flow between reservoirs through the system, preventing 
equilibration. Two exceptions exist, for which the system {is} nevertheless in thermal equilibrium: (i) when all the temperatures $T_i$ are equal, the system is effectively coupled to a single heat reservoir. Therefore, its steady state is an equilibrium state; (ii) When the potential is additively separable in the DoF, namely $U(\mathbf{x})=\sum_iU_i(x_i)$. In this case there is no coupling between the different DoF, and each DoF relaxes to its own equilibrium according to the temperature of the bath it is coupled to (see also discussion on \eqref{eq:equilibrium_condition_multi_bath}).

\subsection*{Entropy production rate}
As discussed above, \eqref{eq:klein_kramers}  can describe a system that relaxes either towards equilibrium or 
into a genuine nonequilibrium steady state. A hallmark of the  difference between these two cases is the total entropy production rate (EPR) in the steady-state, given by \cite{sekimoto_stochastic_2010}
\begin{equation}\label{eq:EPR}
\begin{split}
         \langle\dot{S}\rangle  &= -\sum_i \frac{ \langle \dot{Q}_i \rangle }{T_i} \\
         &= \int \! d\mathbf{x}\, d\mathbf{v}\,\frac{ \xi}{f_{ss}} \sum_i \frac{1}{ T_i} \left[ v_i f_{ss} +  \frac{k_{\rm B} T_i}{m}\frac{\partial f_{ss}}{\partial v_i} \right]^2 \\
         &= \int \! d\mathbf{x}\,d\mathbf{v}\, \frac{\xi}{f_{ss}} \, \left| \mathcal{T}^{-1/2}\cdot \mathbf{v}f_{ss} + \frac{k_{\rm B} \mathcal{T}^{1/2}}{m}\cdot \nabla_{\mathbf{v}}f_{ss}\right|^2 \geq 0
\end{split}
\end{equation}
where $\langle \dot{Q}_i\rangle$ is the average heat flux from the $i$-th reservoir to the system at the steady-state, and $f_{ss} = \lim_{t\to\infty}f_t$ is the steady-state distribution. We note that while both single temperature and additively separable potentials yield $\langle \dot{S} \rangle = 0$ in their steady state, it is \emph{not} immediately obvious from \eqref{eq:EPR} alone that they lead to a vanishing entropy production (the potential does not even appear explicitly in \eqref{eq:EPR}). In this sense, the thermodynamics of the multi-reservoir system is not {particularly intuitive}, and the equilibrium limits are not immediately discernible. We return to this point later when discussing the EPR of the transformed system.

{\subsection*{Strong-damping limit}}\label{sec:overdamped_limit}
In (SI Appendix), we take the strong-damping limit $0 < m/\xi \ll 1$ of \eqref{eq:klein_kramers} and obtain an expression for the steady-state distribution to first order in $\varepsilon = m/\xi$. In this limit, the Liouville flow $\{H, f_{ss}\}$ acts as a small perturbation on the dissipative part, which is an Ornstein-Uhlenbeck process in the velocities (solved by the multi-temperature generalization of Maxwell-Boltzmann distribution). The NESS to leading order reads
\begin{align}
        f_{\rm ss}(\mathbf{x}, \mathbf{v})  =& \left[ \rho(\mathbf{x})- \frac{m}{\xi}\left(\frac{\mathcal{T}^{-1}\nabla_{\mathbf{x}} {U}}{k_{\rm B}}\rho(\mathbf{x}) + \nabla_\mathbf{x}\rho(\mathbf{x})\right)\cdot\mathbf{v} \right]\nonumber\\
        &\times \left(\frac{m}{2\pi k_{\rm B}}\right)^{N/2} |\mathcal{T}|^{-1/2} \exp \left( - \frac{m\mathbf{v}^T\mathcal{T}^{-1}\mathbf{v}}{2k_{\rm B}}\right) \\ &+ \mathcal{O}\left(\left(m/\xi \right)^2\right), \nonumber
\end{align}
where $\rho(\mathbf{x})$ is the solution of the steady-state Smoluchowski equation for a system coupled to multiple reservoirs,
\begin{equation}
    \nabla_{\mathbf{x}}\cdot\left(\frac{\nabla_{\mathbf{x}} {U}}{\xi}\rho(\mathbf{x}) + \frac{k_{\rm B} \mathcal{T}}{\xi}\nabla_\mathbf{x}\rho(\mathbf{x})\right)=0.
\end{equation}
{The normalization of $ f_{\rm ss}(\mathbf{x}, \mathbf{v})$ follows from the normalization of $\rho(\mathbf{x})$ and the fact that the correction is essentially a scalar product with the mean of the centered Gaussian over $\mathcal{T}^{-1/2}\mathbf{v}$ which vanishes by symmetry.}

\section*{Eliminating multiple reservoirs}\label{sec:Transformation}
Systems that are simultaneously coupled to multiple heat reservoirs at different temperatures are ubiquitous and well studied \cite{esposito2012stochastic, seifert2012stochastic, murashita2016overdamped, van2004microscopic, van2006brownian, filliger2007brownian, dotsenko2013two, mancois2018two, bae2021inertial, baldassarri2020engineered,cerasoli2018asymmetry, cerasoli2022spectral, chang2021autonomous, dos2021stationary, pietzonka2018universal, lin2022stochastic, movilla2021energy, abdoli2022tunable, raz2016geometric, chiang2017electrical, ciliberto2013heat, PhysRevE.92.032118, abdoli2025quadrupolar, visco2006work, berut2014energy, berut2016stationary, fogedby2011bound, fogedby2017minimal, leighton2024information, netz2020approach, argun2017experimental, martins2026role, wu2022nonequilibrium}.  For the case of a single particle coupled to different heat baths, experiments typically  use a system analog of Brownian motion, e.g., an electric-circuit in which resistors are held at different temperatures \cite{ciliberto2013heat, chiang2017electrical}, or artificially implement the anisotropic noise acting on a diffusing colloidal particle using an athermal random force \cite{argun2017experimental, berut2014energy, lin2022stochastic, berut2016stationary, martins2026role}. 

The main challenge with multi-temperature systems is that their nonequilibrium dynamics is typically quite complicated. Energy flows through the system create nontrivial circulation \cite{filliger2007brownian, abdoli2025quadrupolar, chang2021autonomous} which is hard to characterize intuitively, and typically requires either solving or simulating the system's dynamics. In what follows, we show how some intuition can nevertheless be gained in such cases.  

\subsection*{Canonical transformation}
A different perspective on such systems (and on their underdamped dynamics in particular) can be obtained by considering the following canonical transformation. We define a generating function of the second kind \cite{LANDAU1976131},
\begin{equation}\label{eq:generating_func}
    G_2(\mathbf P, \mathbf x) = \tilde{T}^{1/2}\,\mathbf P^T \cdot \mathcal T^{-{1}/{2}}\cdot\mathbf x,
\end{equation}
where $\mathbf P$ and $\mathbf x$ are the new momenta and old coordinates, respectively, and $\tilde{T} > 0$ is a scalar parameter with units of temperature. Note that the temperatures $T_i$ of the multiple reservoirs are positive, so $\mathcal{T}$ is positive definite and $G_2$ is well defined.
The new coordinates and momenta in terms of the old ones are given by
\begin{equation}\label{eq:canonical_transformation}
\begin{split}
    \mathbf X &= \tilde{T}^{1/2} \mathcal T^{-{1}/{2}}\mathbf x,\\
    \mathbf P &= \tilde{T}^{-1/2} \mathcal T^{{1}/{2}} \mathbf p,
\end{split}
\end{equation}
and the transformed Hamiltonian is
\begin{equation}\label{eq:transformed_hamiltonian}
    \tilde{H}(\mathbf X, \mathbf P) = \frac{1}{2m}\mathbf P^T (\tilde{T}\mathcal{T}^{-1}) \mathbf{P} + \tilde{U}\left(\mathbf X\right),
\end{equation}
where $\tilde{U}(\mathbf{X})=U(\mathbf{x})$ is the potential in terms of the transformed coordinates\footnote{Throughout the rest of the manuscript, lower case variables ($\mathbf{x}$, $\mathbf{v}$) refer to the original coordinates, and upper case variables ($\mathbf{X}$, $\mathbf{V}$) and tilde-annotated letters ($\tilde{U}, \tilde{T}, \tilde{\mathbf{F}}$, etc.) are used for the transformed description.}. Notably, the transformed kinetic energy is anisotropic, {and as a consequence the velocity is $\bm{V}\equiv \dot{\bm{X}}=\frac{\partial \tilde{H}}{\partial \bm{P}} = \tilde{T}\mathcal{T}^{-1}\bm{P}/m$ (and not $\bm{P}/m$ as in the isotropic case)}. We elaborate further on this important feature and describe its physical significance below, but first we apply the transformation to \eqref{eq:langevin}-\eqref{eq:klein_kramers} to obtain a description of the stochastic evolution in these new coordinates.

\begin{figure}
    \centering
    \includegraphics[width=.8\linewidth]{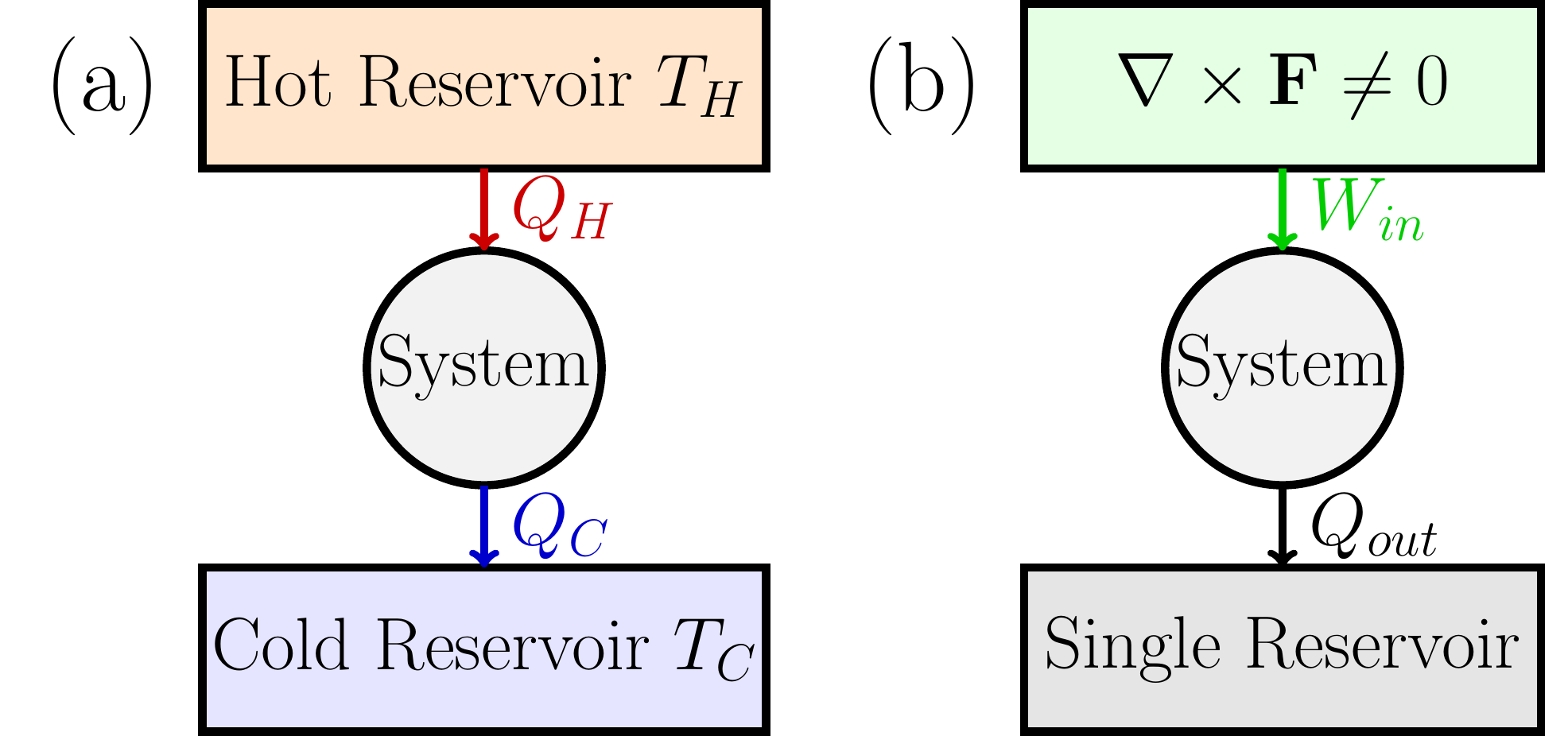}
        \caption{Schematic comparison of the original and transformed systems. (a) The original, multi-reservoir system (illustrated for two, but the formalism is general and applies to any number of reservoirs), with heat flowing between the reservoirs through the system. (b) The transformed, single reservoir system, is driven out of equilibrium by a non-conservative curl-force that performs work on it. The work is dissipated into the single reservoir as heat.}
    \label{fig:transformation}
\end{figure}

The Langevin equation becomes
\begin{equation}
    \label{eq:langevin_transformed}
    \begin{split}
    d\mathbf{X}(t)&=\mathbf{V}(t)dt\\
    m d\mathbf{V}(t)&=-\tilde{T}\mathcal{T}^{-1}\cdot\nabla_\mathbf{X} \tilde{U}(\mathbf{X})|_{\mathbf{X}=\mathbf{X}(t)}dt - \xi\mathbf{V}(t)dt \\
    &\qquad+ \sqrt{2\xi k_{\rm B} \tilde{T}} d\mathbf{W}(t)
    \end{split}
\end{equation}
where the transformation of $\mathbf{v} \to \mathbf{V}$ has eliminated the $\mathcal{T}^{1/2}$ matrix multiplying the white noise $\mathbf{W}(t)$, such that the transformed system describes an underdamped process with coupling to a single bath with temperature $\tilde{T}$ and having the same drag coefficient $\xi$ as of the original system. We schematically illustrate the effect of the transformation in Fig.~\ref{fig:transformation}.

Similarly, by applying the same transformation to the Klein-Kramers equation we obtain 
\begin{equation}    \label{eq:klein_kramers_transformed}
\begin{split}
  \frac{\partial \tilde{f}_t(\mathbf{X}, \mathbf{V})}{\partial t} =& \frac{(\tilde T \mathcal T^{-1}\cdot\nabla_{\mathbf{X}} \tilde{U}) \cdot \nabla_{\mathbf{V}}\tilde{f}_t}{m}-\nabla_{\mathbf{X}} \tilde{f}_t \cdot \mathbf{V} \\ &\qquad+ \frac{\xi}{m} \nabla_{\mathbf V} \cdot\left(\mathbf{V} \tilde{f}_t + \frac{k_{\rm B} \tilde T}{m} \nabla_{\mathbf{V}}\tilde{f}_t\right)\\
  =& \{\tilde{H}, \tilde{f}_t \} + \frac{\xi}{m} \nabla_{\mathbf V} \cdot\left(\mathbf{V} \tilde{f}_t + \frac{k_{\rm B} \tilde T}{m} \nabla_{\mathbf{V}}\tilde{f}_t\right),
  \end{split}
\end{equation}
which has a scalar diffusion coefficient $D=\frac{\xi k_{\rm B} \tilde{T}}{m^2}$ in place of the original system's diffusion matrix with multiple temperatures.

The last line in \eqref{eq:klein_kramers_transformed} suggests that in the new coordinates the system is an equilibrium system, since $\tilde H$ is a genuine Hamiltonian and all DoF are coupled to a single temperature heat bath. As we discuss below, this is \emph{not} the case, although the Hamiltonian structure is nevertheless useful and can provide intuition on the  the system's behavior. To this end, we first address the aforementioned non-standard structure of the transformed Hamiltonian, the force $-\tilde T \mathcal T^{-1}\nabla_{\mathbf{X}} U$ appearing in the drift part of the transformed equation, followed by a discussion on the nonequilibrium nature of the transformed system.

\subsection*{Hamiltonian curl-forces}
\label{sec:hamiltonian_curl_forces}
The Hamiltonian obtained in \eqref{eq:transformed_hamiltonian} is of a non-standard form: it contains the anisotropic quadratic form $\frac{1}{2m}\mathbf{P}^T\mathcal{A}\mathbf{P}$, with $\mathcal{A} \equiv \tilde{T}\mathcal{T}^{-1}$ a symmetric positive definite matrix. Such Hamiltonians, with a general anisotropic kinetic energy term, were studied by Berry and Shukla \cite{berry2015hamiltonian}, who showed that they generically produce a velocity-independent but non-conservative force --- a curl force. We briefly recapitulate 
such systems (for a more detailed review of curl-forces and their dynamical properties, see also Refs.~\cite{berry2012classical,berry2016curl,berry2023quantum,berry2024quantising,berry2025six}). For simplicity, we first discuss systems with two DoF, and then extend to systems with more DoF. The following discussion follows closely Ref. \cite{berry2015hamiltonian}.

\subsubsection*{Deterministic dynamics in two dimensions}
The standard Hamiltonian of a classical deterministic system in dimension 2 in the absence of a magnetic field has of the form in \eqref{Hamiltonian_0} 
with $\mathbf p=(p_{1},p_{2})$ being now the canonical momentum conjugate to the cartesian coordinates $\mathbf x = (x_1,x_2)$\footnote{In principle, one can use any set of canonical coordinates and momenta, not necessarily cartesian coordinates. However, in the non cartesian case it is less trivial to identify whether the force field is velocity independent. Therefore, we assume here cartesian coordinates for simplicity.}, and $m$ is the particle's mass. The equations of motion read 
\begin{eqnarray} \label{eq:EqOfMotion}
    \dot x_i = \frac{\partial H}{\partial p_i} = \frac{p_i}{m}, \qquad 
    \dot p_i=-\frac{\partial H}{\partial x_i} = -\frac{\partial {U}(\mathbf x)}{\partial x_i},
\end{eqnarray}
and the force is defined as 
\begin{equation} \label{eq:NewtonForce}
    \mathbf F (\mathbf{x})\equiv m\ddot{\mathbf x} = -\nabla {U}(\mathbf x),
\end{equation}
which is conservative, as it is a gradient of the potential. 

One can, in principle, consider Hamiltonians that are not of the form \eqref{Hamiltonian_0}, whereby the force is still defined by Newton's second law as $\mathbf F \equiv m \ddot{\mathbf x}$ but in general may be velocity dependent and not a function of position alone. Berry and Shukla showed \cite{berry2015hamiltonian} that under {some} assumptions, the general Hamiltonian structure that leads to a velocity independent force field {(see Ref.~\cite{yip2024general} for the most general form)}, is 
\begin{equation}
    \tilde H(\mathbf P,\mathbf X) = \frac{1}{2m}\mathbf P^T \mathcal{A} \mathbf P + \tilde{U}(\mathbf X),
\end{equation}
where $\mathcal{A}$ is a symmetric, invertible, and position independent matrix. In this case,
\begin{equation}
\tilde{\mathbf F} \equiv m\ddot{\mathbf{X}} = -\mathcal{A}\nabla_{\mathbf{X}} \tilde{U}(\mathbf X).
\label{eq:HCF-def}
\end{equation}
Without loss of generality, we work in the (cartesian) basis where $\mathcal{A}$ is diagonal, 
\begin{eqnarray}
    \mathcal{A} = \begin{pmatrix}
         \alpha & 0 \\
         0 & \gamma
    \end{pmatrix} ,
\end{eqnarray}
where $\alpha$ and $\gamma$ are unitless constants.
The curl of the force field in a system with 2 DoF is a scalar, given by
\begin{equation}\label{eq:berry_curl}
    \nabla\times\tilde{\mathbf F} = \left( \alpha - \gamma \right) \partial_{X_1}\partial_{X_2}\tilde{U}. 
\end{equation}
Therefore, the force field is not curl free, unless either $\alpha=\gamma$ or $\tilde{U}(\mathbf X)$ is everywhere additively separable in the coordinates $\mathbf X$.
    
\subsubsection*{Deterministic dynamics in higher dimensions}\label{sec:high_d_generalization}
The above argument for two DoF immediately generalizes to 
more DoF as follows: instead of $\nabla \times\tilde{\mathbf{F}}$, consider the matrix $(\nabla \tilde{\mathbf{F}})_{ij}\equiv \partial_{{j}}\tilde{F}_{{i}}$. From \eqref{eq:HCF-def} it follows that $\nabla \tilde{\mathbf{F}} = -\mathcal{A}\nabla\nabla \tilde{U}(\mathbf X)$ where $(\nabla\nabla \tilde{U})_{i,j} = \partial_{X_i}\partial_{X_j}\tilde{U}$, namely the Hessian of $\tilde{U}$. If $\nabla \tilde{\mathbf{F}}$ is symmetric, then by Poincaré's lemma $\tilde{\mathbf{F}} = -\nabla \phi (\mathbf X)$ for some scalar potential $\phi(\mathbf X)$ (we assume a simply connected manifold). Therefore, $\tilde{\mathbf{F}}$ is not a gradient of a potential only if $\mathcal{A}\nabla\nabla \tilde{U}\neq \nabla\nabla \tilde{U} \mathcal{A}$ (note that both the Hessian and $\mathcal{A}$ are symmetric matrices and the product of symmetric matrices is generally not symmetric). This means that the force field $\tilde{\mathbf{F}}$ is not conservative --- it is not a gradient of a potential --- when the matrix $\mathcal{A}$ and the Hessian of $\tilde{U}$ do not commute, namely $[\mathcal{A},\nabla\nabla \tilde{U}(\mathbf{X})]\neq 0$ at some $\mathbf{X}$.  

\subsubsection*{{Heat exchange in the multi-reservoir description, revisited}}\label{sec:separable_coordinates}
We now transform the condition for the generalized curl of the force to vanish back to the multi-reservoir description. The condition for a \emph{vanishing curl-force} becomes
\begin{equation}\label{eq:equilibrium_condition_multi_bath}
    [\mathcal{T}, \nabla_\mathbf{x}\nabla_\mathbf{x}U(\mathbf{x})] = 0, \qquad \forall\mathbf{x}.
\end{equation}
Component-wise, this reads $(T_i - T_j)\partial_{x_i}\partial_{x_j}U(\mathbf{x}) = 0$, which is  satisfied if and only if the potential is additively separable for any DoF coupled to a bath with a different temperature (DoF $i\neq j$ with $T_i=T_j$ may have a non-vanishing $\partial_{x_i}\partial_{x_j}U(\mathbf{x})$). In this case, the DoF are decoupled and heat cannot flow through the system, so each DoF separately equilibrates with its own bath. Applying a non-conservative force in the single-reservoir description is therefore analogous to allowing the multi-reservoirs to exchange heat in the original description.

\subsection*{Stochastic dynamics with Hamiltonian curl-forces}
\subsubsection*{Hamiltonian, yet nonequilibrium dynamics}
Having identified the transformed Hamiltonian as a curl-force Hamiltonian with positive definite anisotropy $\mathcal{A}=\tilde{T}\mathcal{T}^{-1}$ and correspondingly the curl-force $-\tilde{T}\mathcal{T}^{-1}\nabla_{\mathbf{X}}\tilde{U}$, we now show how this is manifested in the nonequilibrium nature of the transformed system. First, we note that \eqref{eq:klein_kramers_transformed} seems to suggest that the Boltzmann-Gibbs distribution $\tilde{f}_{\rm eq}\sim \exp(-\tilde{H}/k_{\rm B}\tilde{T})$ associated with the curl-force Hamiltonian $\tilde{H}$ is the steady-state distribution of the transformed system --- that is, the system in fact is in thermal equilibrium with a single heat bath. However, direct substitution of $\tilde{f}_{\rm eq}$ shows that this is \emph{not} the case. Importantly, the reason lies not in $\tilde H$ itself, which generates genuine Hamiltonian dynamics (albeit the corresponding conserved quantity is not the regular classical energy), but in the manner the system is coupled to the bath.

\subsubsection*{Origin of nonequilibrium dynamics}
The standard derivation of the Boltzmann distribution in the canonical ensemble assumes very little about the system, the thermal bath, and the interaction between them {beyond that the latter is weak and the bath is large}. Moreover, it does not assume a specific structure for the Hamiltonian. Nevertheless, the Boltzmann distribution  $p^{eq}\sim e^{\tilde H/k_B\tilde T}$ is not the steady state of the system with the curl-force Hamiltonian, even though its dynamics in the absence of coupling to the thermal bath follows Hamiltonian equations of motion. 
The reason is that even though the system's Hamiltonian is conserved when decoupled from the bath, the total Hamiltonian of the system and the bath together is not conserved.

The Klein-Kramers equations \eqref{eq:klein_kramers_transformed} are a coarse-grained description of the full system and bath dynamics under proper assumptions, e.g. large separation between the bath and system relaxation timescales, infinitely large  bath with mixing DoF, weak coupling between the bath and the system \cite{rey2006open}. {The origin of a non-Boltzmann-Gibbs stationary distribution may be rationalized in different ways. In the deterministic setting that includes both the system and the thermal bath, the total Hamiltonian is given by
\begin{equation}\label{eq:total_energy}
\begin{split}
    H_{\rm tot} &= H_{\rm sys}(\mathbf P_{\rm sys},\mathbf X_{\rm sys})+H_{\rm bath}(\mathbf P_{\rm bath},\mathbf X_{\rm bath}) \\ &\qquad +  U_{\rm int}(\mathbf X_{\rm sys},\mathbf X_{\rm bath}).
\end{split}
\end{equation}
where $H_{\rm sys}(\mathbf P_{\rm sys},\mathbf X_{\rm sys})$ is the Hamiltonian of the system had it not been coupled to the bath, $H_{\rm bath}(\mathbf P_{\rm bath},\mathbf X_{\rm bath})$ is the Hamiltonian of the bath DoF, and $U_{\rm int}(\mathbf X_{\rm sys},\mathbf X_{\rm bath})$ is the interaction energy between the bath and system. Due to the non-standard kinetic energy of the system, $\frac{1}{2m}\mathbf{P}^T\mathcal{A}\mathbf{P}$, the instantaneous force acting on the system is $ \mathbf{F}_{\rm sys} = -\mathcal{A}\cdot\nabla U_{\rm sys}(\mathbf X_{\rm sys})
    - \nabla U_{\rm int}(\mathbf X_{\rm sys},\mathbf X_{\rm bath})$, while the force in a truly Hamiltonian setting with $H_{\rm tot}$ in \eqref{eq:total_energy} would read $\mathbf{F}_{\rm sys} = -\mathcal{A}\cdot\nabla [U_{\rm sys}(\mathbf X_{\rm sys}) + U_{\rm int}(\mathbf X_{\rm sys},\mathbf X_{\rm bath})]$. Therefore, the system-bath interaction force is not the Hamiltonian force, and the total Hamiltonian is not conserved. 
}

As a final remark, we mention that the drag force  $-\xi \mathbf{V}$  we obtained {in \eqref{eq:langevin_transformed}} through the canonical transformation is indeed the physically-consistent one for a system with a non-standard Hamiltonian. The canonical momentum of the curl-force system is not by itself a physically measurable quantity, so the Stokes friction should be proportional to the immersed particle's velocity.
This ensures that when the {curl force} is turned off, 
the particle diffuses like a regular particle, and is in line with the literature on Brownian motion in other cases where the canonical momentum $\mathbf{P}$ does not coincide with the inertial momentum $m \mathbf{V}$, for example for a charged particle in a magnetic field \cite{taylor1961diffusion, kurcsunoglu1963brownian, czopnik2001brownian, simoes2005kramers, lisy2013brownian, chun2018emergence, abdoli2020nondiffusive}.

\subsubsection*{{Invariance of entropy production rate under transformation}}\label{sec:transformed_EPR}
The derivation of the entropy production rate (EPR) \cite{sekimoto_stochastic_2010} in \eqref{eq:EPR} is based on the fact that in the steady-state, the system's entropy is constant, such that the \emph{total} EPR is equal to that produced in the reservoirs (a single reservoir in the {transformed setting}), and is quantified by the heat exchanged with the system. To this end, the EPR is indifferent to the mechanism driving the system out of equilibrium, i.e.\ coupling to multiple reservoirs or applying a curl-force. Therefore, \eqref{eq:EPR} readily applies the transformed system. In terms of the new coordinates and bath temperature $\tilde{T}$, it reads
\begin{equation} \label{eq:EPR_transformed}
    \langle \dot{\tilde{S}} \rangle = \int \! d\mathbf{X}\, d\mathbf{V}\, \frac{\xi}{\tilde{f}_s \tilde{T}}\left(\mathbf{V}\tilde{f}_s + \frac{k_{\rm B} \tilde T}{m}\nabla_\mathbf{V}\tilde{f}_s\right)^2.
\end{equation}
We further note that $\langle \dot{S}\rangle = \langle \dot{\tilde{S}}\rangle$, as is evident by direct substitution of the transformation \eqref{eq:canonical_transformation} in \eqref{eq:EPR}, or more generally considering the covariant form of the stochastic dynamics \cite{graham1977covariant, polettini2013generally, miller2025covariant}, where the entropy production rate is found to be a scalar. 
As in the multi-reservoir case, \eqref{eq:EPR_transformed} does \emph{not} explicitly depend on the curl-forces, and substitution of $\tilde{f}_s$ is required in order to determine whether the EPR is vanishing or not.

However, the transformed representation allows for another way to compute the EPR. To explicitly determine the role of the curl-force $\tilde{\mathbf{F}}=-\tilde{T}\mathcal{T}^{-1}\nabla_\mathbf{x} U$ in generating entropy, we note that in a NESS, the energy balance on the system must vanish, such that the work applied on the system, the heat released to the bath, and the entropy production rate relate through
\begin{equation} \label{eq:steady_state_relation}
    \langle \dot{\tilde W} \rangle = -\langle \dot{\tilde Q} \rangle = \tilde{T} \langle \dot{\tilde{S}}\rangle.
\end{equation}
This relation is only valid in the single-reservoir description, as in the multi-reservoir case the energy balance is expressed in terms of the sum of heat contributions exchanged with each bath at the respective temperatures, and work and heat can be distributed between the different reservoirs.
The rate of work (i.e., power) applied on the system is
\begin{equation}\label{eq:work_curl}
\begin{split}
       \langle \dot{\tilde W} \rangle &= \int \! d\mathbf{X}d\mathbf{V}\tilde{f}_s(\bX,\bV)[\mathbf{\tilde F}\cdot\mathbf{V}]   \\&= -\tilde{T}\int \! d\bX d\bV\tilde{f}_s(\bX,\bV)\,[\mathcal{T}^{-1}\nabla_{\mathbf{X}} \tilde{U}\cdot\mathbf{V}],
\end{split}
\end{equation}
where integrating \eqref{eq:klein_kramers_transformed} over $\mathbf{V}$ in the steady state gives $0 = \nabla_\mathbf{X}\cdot\int\mathbf{V}f_sd\mathbf{V}$, such that the gradient part of the force $-\tilde{T}\mathcal{T}^{-1} \nabla \tilde{U}$ vanishes after integrating \eqref{eq:work_curl} by parts. The work, and consequently the entropy production rate, therefore depend exclusively on the rotational (non-gradient) part of the force field. This trivial statement is exemplified 
in the entropy production rate for the exactly solvable Ornstein-Uhlenbeck process (see applications below), where the role of the curl-force in driving the system out of equilibrium is particularly obvious. 

\section*{Applications}\label{sec:applications}
We now analyze two examples for systems 
driven out of equilibrium via coupling 
to reservoirs at different temperatures within the context of 
their curl-force representations. Additionally, we analyze the role of the curl of the force field in a general nonequilibrium underdamped Ornstein–Uhlenbeck (OU) process in $N$-dimensions, and show explicitly that it is the source of any nonequilibrium current and in turn of the entropy production. Finally, we use the Hamiltonian curl-force framework to find the stall-forces required to equilibrate a multi-reservoir system.

\subsection*{Brownian gyrator in dimension 2} \label{sec:brownian_gyrator}
The Brownian gyrator is one of the simplest models for a stochastic heat engine. Originally proposed by Filliger and Reimann \cite{filliger2007brownian} and further studied extensively \cite{dotsenko2013two, mancois2018two, bae2021inertial, baldassarri2020engineered, cerasoli2018asymmetry, cerasoli2022spectral, argun2017experimental, martins2026role, chang2021autonomous, dos2021stationary, movilla2021energy, pietzonka2018universal, lin2022stochastic, abdoli2022tunable, chiang2017electrical}, it describes a Brownian particle diffusing in the presence of an anisotropic harmonic potential,
\begin{equation}\label{eq:gyrator_potential}{U}(x, y) = \frac{1}{2}\left( kx^2 + 2uxy + ky^2 \right),\end{equation}
with $|u|<k$, and two different random noises amplitudes in the $x$ and $y$ directions corresponding to temperatures $T_x$ and $T_y$, respectively.  
Due to the harmonicity of the potential, the  stochastic process is an OU process, which is exactly solvable \cite{risken1996fokker}. When $T_x\ne T_y$ the nonequilibrium steady-state of this system exhibits rotational motion, whose magnitude is determined by both, the anisotropy of the potential, determined by $u/k$, and the temperature difference between the two heat baths, $T_y-T_x$ \cite{mancois2018two, bae2021inertial}. By applying the transformation of \eqref{eq:canonical_transformation} 
we uncover an alternative interpretation for the source of rotational motion: it can be viewed as a result of a \emph{deterministic} curl-force acting on a particle coupled to a single heat bath. The corresponding potential in the curl-force representation is
\begin{equation}\label{eq:gyrator_curl_potential}
    \tilde{U}(X, Y) = \frac{1}{2\tilde{T}}\left( {kT_x}X^2+2{u\sqrt{T_x T_y}XY}+kT_y Y^2 \right),
\end{equation}
and the total force due to the potential is \begin{equation}
\begin{split}
    \tilde{\mathbf{F}}(X, Y) &\equiv -{\tilde{T}\mathcal{T}^{-1}}\cdot\nabla_\mathbf{X} \tilde{U} \\ &=-(kX+\sqrt{\frac{T_y}{T_x}} uY)\hat{X}-(kY+\sqrt{\frac{T_x}{T_y}} uX)\hat{Y}. 
\end{split}
\end{equation} The curl of the force field in this case is constant, given by 
$\nabla\times\tilde{\mathbf{F}}=u\left(\frac{T_y - T_x}{\sqrt{T_x T_y}}\right)$.
It is non-vanishing unless either $T_x = T_y$ (i.e.\ a single thermal bath), or $u=0$ for which 
$U$, 
and therefore $\tilde U$, are additively separable since the coordinates are decoupled. 
For $T_x>T_y$ and $u>0$, the curl is negative, which corresponds to a force field with a clockwise circulation and is consistent with the results obtained for the inertial Brownian gyrator \cite{mancois2018two, bae2021inertial}. Notably, this qualitative result is purely mechanical 
and does \emph{not} require a solution of the Klein-Kramers equation governing the system's stochastic dynamics. 
It thus provides a simple and intuitive means to understanding the origin, magnitude, and direction of the Brownian particle's rotation in the two-temperature setup.

It is further instructive to compare the energetics of the two descriptions. In the steady state 
heat $\langle \dot{Q}_x \rangle = -\langle \dot{Q}_y \rangle$ flows from the hot to the cold reservoir through the system, and work is neither applied nor extracted unless an additional external non-conservative force is applied. In the transformed coordinates, the picture is slightly different, as there is only a single heat bath. To find the heat and work rates, we require the steady state distribution, which can be obtained 
directly by solving the Klein-Kramers equation for the curl-force OU process \cite{risken1996fokker}, or by transforming the known solution of the underdamped Brownian gyrator \cite{mancois2018two, bae2021inertial} to the curl-force coordinates. An explicit solution is provided in (SI Appendix). The solution is a Gaussian $\tilde{f}_s(\mathbf{Z}) \sim \exp(-\frac{1}{2}\mathbf{Z}^T \tilde{C}^{-1} \mathbf{Z})$, where $\mathbf{Z} = (X, Y, V_X, V_Y)^T$ and $\tilde C$ is the steady-state covariance matrix. 

The energetics of the system follow directly from $\tilde{C}$. The power applied on the system by the non-conservative force is
\begin{equation}\label{eq:work}
    \langle \dot{\tilde{W}} \rangle = \langle \tilde{\mathbf{F}} \cdot {\mathbf{V}} \rangle = 
\frac{k_{\rm B} \tilde{T} \xi u^2(T_y-T_x)^2 }{2T_x T_y (\xi^{2}  k + u^{2}  m)},
\end{equation}
where we note that $\langle \dot{\tilde{W}}\rangle \propto |\nabla \times \tilde{\mathbf{F}}|^2$, such that the curl of the force field is a measure of how far the system is from thermal equilibrium, in the sense that a larger curl implies that more work is applied on the system. 
\eqref{eq:work} also implies $\langle \dot{\tilde W}\rangle\geq0$: positive work is always applied on the system regardless of the direction of gyration (determined by whether $T_x > T_y$ or $T_x < T_y$), which is a manifestation of the second law: work cannot be extracted from a single heat bath. This highlights the difference of this transformed description from the corresponding two-baths description, where heat flows from the hot to the cold bath \cite{bae2021inertial}, and the direction of the flow can be reversed by interchanging the bath temperatures.

Given the solution for $\tilde C$, the average heat dissipation rate is also computed directly as $\langle \dot{\tilde{Q}} \rangle= -\frac{2\xi}{m}\left(\frac{m}{2}(\langle V_X^2\rangle + \langle V_Y^2\rangle) - k_{\rm B} \tilde{T}\right)$ \cite{sekimoto_stochastic_2010}. Since  the system is in a steady state, the work done by the non-conservative force is fully dissipated into the bath as heat. It then follows from \eqref{eq:steady_state_relation} that the entropy production rate is
\begin{equation}
\begin{split}
        \langle \dot{\tilde S}\rangle&=\langle \dot{\tilde S}\rangle_\text{system}+\langle \dot{\tilde S}\rangle_\text{bath}=\langle \dot{\tilde S} \rangle_\text{bath}=-\frac{\langle \dot{\tilde Q}\rangle}{\tilde T} \\ &= \frac{k_{\rm B} \xi u^2(T_y-T_x)^2 }{2T_x T_y (\xi^{2}  k + u^{2}  m)} \geq 0,
\end{split}
\end{equation}
which is independent of $\tilde{T}$ and is identical to the EPR obtained for the underdamped gyrator \cite{bae2021inertial}, which serves as a sanity check for the correctness of our results. As mentioned above, this follows by consistency since the two descriptions are related via a coordinate transformation \cite{graham1977covariant,polettini2013generally, miller2025covariant}. This stands in contrast to the work and heat, which depend on the temperature $\tilde{T}$ in the curl-force system.

\subsubsection*{Conditions for equal energy flows}
A choice of $\tilde{T}$ that can be of particular interest is to require that $\langle \dot{\tilde{W}}\rangle = \langle \dot Q _X \rangle$ and $\langle \dot{\tilde{Q}}\rangle = \langle \dot{Q}_Y\rangle$ (for $T_x>T_y$), such that the energy flows into or out of the system is equal in both descriptions. An exact solution can be obtained by noting that the entropy production rate is invariant under the coordinate transformation. We can compare its expressions in both representations,
\begin{equation}\label{eq:compare_entropies}
    \langle \dot{S} \rangle = -\left(\frac{\langle\dot{Q}_x\rangle}{T_x} + \frac{\langle\dot{Q}_y\rangle}{T_y}\right) = -\frac{\langle \dot{\tilde{Q}}\rangle}{\tilde{T}} = \langle \dot{\tilde{S}}\rangle.
\end{equation}
Plugging in the condition $\langle \tilde{W} \rangle = |\langle \dot{\tilde{Q}}\rangle| = |\langle\dot{Q}_x\rangle| = |\langle\dot{Q}_y\rangle|$ along with $\langle Q_x\rangle = -\langle Q_y\rangle$, and re-arranging, we obtain
\begin{align}\label{eq:equal_flow_temp}
    \tilde{T} =  {\left|T_x^{-1} - T_y^{-1}\right|^{-1}}, \qquad T_x \neq T_y.
\end{align}
Notably, if $T_x \gg T_y$ (or $T_y \gg T_x$) the bath temperature in the transformed system, for which the heat flows are equal 
in both representations, is approximated by the temperature of the colder bath.
We note that this result is general for two-reservoir systems with any potential and not only in the harmonic case.

\subsection*{Gyrator in an azimuthal-symmetric potential }\label{sec:ring_gyrator}
\begin{figure*}[t!]
    \centering
\includegraphics{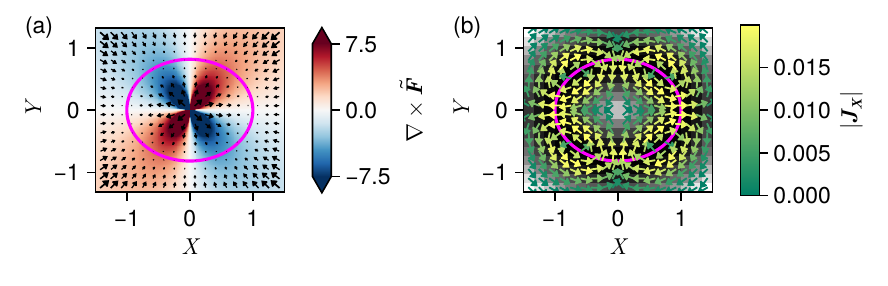}
    \caption{
    (a) The force field $\mathbf{\tilde{F}}=m\ddot{\mathbf{X}}$ of the single-bath system dual to the ring gyrator (arrows), and the value of its curl $\nabla\times\mathbf{\tilde{F}}$ (color), for $k = 20, r_0 = 1, T_x = 1, T_y = 3/2$. Unlike the case of the harmonically confined Brownian gyrator, here $\nabla\times\mathbf{\tilde{F}}$ is not constant, and has a quadrupolar structure instead.
    (b) Simulation results for the probability density and currents averaged over 25,000 Langevin trajectories of the same system coupled to a bath with temperature $k_{\rm B} \tilde{T}= 1$ and drag coefficient $\xi=1$.
    The steady state probability density function is shown in the background (white - low density, black - high density). The arrows indicate the empirical spatial probability current $\mathbf{J}_\mathbf{X}=\langle \mathbf{V} | \mathbf{X}\rangle f$, where $\langle \mathbf{V} | \mathbf{X}\rangle$ is the average velocity conditioned on position.
    A positive value of $\nabla\times\mathbf{\tilde{F}}$ in (a) corresponds to counter-clockwise rotation in (b) and a negative value corresponds to clockwise rotation. These directions are in agreement with those found for the gyrator in this potential \cite{abdoli2025quadrupolar} for $T_y>T_x$.
    The ring $x^2 + y^2 = r_0^2$, which is transformed into an ellipse $T_x X^2  + T_y Y^2 = \tilde{T}r_0^2$ in the dual system, is shown in a solid pink line in both figures.}
    \label{fig:curl_ring_gyrator}
\end{figure*}

To illustrate the usefulness of this framework in a system where an exact solution is not known, we consider a particle in 2D system, elastically coupled to the origin by a {Hookean potential ${U}(\mathbf{x}) = \frac{1}{2}k(|\mathbf{x}|-r_0)^2$ where $r_0$ is the resting length of the elastic} \cite{abdoli2025quadrupolar}. {Note that the potential is \emph{not} harmonic in $\mathbf{x}$.} The system is coupled to two orthogonal sources of thermal noise with temperatures $T_{x}$ and $T_{y}$, as in the case of the harmonically confined gyrator. Remarkably, the probability flux in this system has been shown to have a quadrupolar structure, with four alternating regions of clockwise and counter-clockwise rotating currents \cite{abdoli2025quadrupolar}. Transforming to the curl-force representation we obtain the force field
\begin{equation}
\begin{split}
       \tilde{\mathbf{F}}(\mathbf{X}) &\equiv -\tilde{T}\mathcal{T}^{-1}\cdot\nabla_{\mathbf{X}}\tilde{U} \\ &= -k\left( 1 - \frac{\sqrt{\tilde{T}}}{\sqrt{T_x{X^2} + T_y {Y^2}}}r_0 \right) \mathbf{X},
\end{split}
\end{equation}
which is radial and is concentrated around a ring, which in the dual system is the ellipse $T_x X^2 + T_y Y^2 = \tilde{T} r_0^2$. Fig.~\ref{fig:curl_ring_gyrator}a shows the force field and the corresponding curl,
\begin{equation}
    (\nabla \times \tilde{\mathbf{F}})\cdot\hat{Z} = \frac{r_0 k {\tilde T}^{1/2} (T_y - T_x)XY}{(T_x X^2 + T_y Y^2)^{3/2}}.
\end{equation}
Unlike for the harmonic gyrator, the curl of the force field is \emph{not} constant and has a quadrupolar structure. For $T_x>T_y$ ($T_y > T_x$), it is negative (positive) in the 1st and 3rd (2nd, 4th) quadrants, corresponding to clockwise (counter-clockwise) circulation. As before, the direction of circulation predicted by the sign of $\nabla \times \tilde{\mathbf{F}}$ coincides with 
the results in \cite{abdoli2025quadrupolar}. We numerically verify that the same direction is also observed in the curl-force system by performing Langevin simulations \cite{rackauckas2017differentialequations, rackauckas2017adaptive, rossler2010runge} of the dynamics, as shown in Fig.~\ref{fig:curl_ring_gyrator}b. Unlike the harmonic Brownian gyrator, in this case the steady-state solution has no exact analytical solution and therefore requires approximations or numerical solutions. Hence, 
the easy exact computation of $\nabla \times \tilde{\mathbf{F}}$ provides a simple method for gaining \emph{qualitative insight} about the system that is otherwise less straightforward to obtain.

\subsection*{Harmonic potential in N-dimensions: role of the generalized curl}\label{sec:N_dim_exact_sol}
The results for the (harmonic) Brownian gyrator in two dimensions can be generalized to an arbitrary confining harmonic potential in $N$-dimensions. In (SI Appendix), we derive an expression for the steady-state distribution for an underdamped OU process with force field $\tilde{\mathbf{F}}=-\mathcal{A}\cdot\nabla\nabla \tilde{U}\cdot\mathbf{X}\equiv-\tilde{\Phi}\mathbf{X}$. 
Utilizing the structure of the underdamped process, the problem is simpler than solving the arbitrary $N$-dimensional OU process in phase space \cite{risken1996fokker}. A steady-state solution exists if $\tilde{\mathbf{F}}$ is confining \cite{risken1996fokker}. We show that in this case the blocks of the steady-state covariance matrix
\begin{equation}
    \tilde{C} = \begin{pmatrix}
        \tilde{C}_{XX} & \tilde{C}_{XV} \\
        \tilde{C}_{XV}^T & \tilde{C}_{VV}
    \end{pmatrix}
\end{equation}
can be written in terms of the equilibrium solution plus a correction due to the nonequilibrium circulation $\tilde{C}_{XV}$,
\begin{subequations}\label{eq:equilibrium_correction}
    \begin{align}
    \tilde{C}_{VV} &= \frac{k_{B}\tilde{T}}{m}\,\mathbb{I} - \frac{1}{2\xi}\left(\tilde{\Phi} \tilde{C}_{XV} - \tilde{C}_{XV}\tilde{\Phi}^{T}\right), \\
    \tilde{C}_{XX} &= (m\,\tilde{C}_{VV} - \xi \tilde{C}_{XV})(\tilde{\Phi}^{-1})^T,
    \end{align}
\end{subequations}
where $\tilde{C}_{XV}$ is an antisymmetric matrix determined by solving the $N(N-1)/2$ linear equations
\begin{equation}\label{eq:covariance_source}
\begin{split}
    2\xi^{2} &\left(\tilde{\Phi} \tilde{C}_{XV} + \tilde{C}_{XV}\tilde{\Phi}^{T}\right) \\
    &+ m\left(\tilde{\Phi}^{2}\tilde{C}_{XV} - 2\tilde{\Phi} \tilde{C}_{XV}\tilde{\Phi}^{T} + \tilde{C}_{XV}(\tilde{\Phi}^{T})^{2}\right)
    \\ &\qquad = 2 k_{\rm B} \tilde{T} \xi (\tilde{\Phi}-\tilde{\Phi}^T) \\ &\qquad = 2k_{B}\tilde{T}\xi[\mathcal{A},\nabla\nabla \tilde{U}].
\end{split}
\end{equation}
An exact solution for $\tilde{C}_{XV}$ is then obtained by a spectral decomposition of $\tilde{\Phi}$ (SI Appendix).
The form of \eqref{eq:covariance_source} highlights the role of the generalized curl of the force field, defined by the commutator $[\mathcal{A}, \nabla\nabla \tilde{U}]$, or equivalently, by (twice) the asymmetric part of the force matrix $\tilde{\Phi}-\tilde{\Phi}^T$. A vanishing curl implies $\tilde{C}_{XV}=0$, and as a result $\tilde{C}_{XX}$ and $\tilde{C}_{VV}$ reduce to their equilibrium values. 
The same is also reflected in the entropy production rate, which can be written compactly as
\begin{equation}\label{eq:OU_EPR_result}
    \langle\dot{S}\rangle = -\frac{1}{2\tilde{T}}\mathrm{Tr}\left([\mathcal{A},\nabla\nabla \tilde{U}]\,\tilde{C}_{XV}\right),
\end{equation}
thus highlighting the role of the curl of the force on entropy production. While it is already well known that the asymmetric part of the drift is responsible for driving the system out of thermal equilibrium \cite{kwon2005structure, wu2021landscape, tomita1974irreversible, godreche2019characterising, dieball2022coarse, mazzolo2023nonequilibrium, miller2025covariant, netz2020approach}, we find it instructive to present these results here in terms of the curl, thus demonstrating how Hamiltonian curl-forces allow for a compact way of describing non-conservative force fields and highlighting their thermodynamic significance.

These results may be directly transformed to the multi-bath description by noting that $\tilde{C}_{XV}= \langle \mathbf{X}\mathbf{V}^T\rangle = \mathcal{A}^{1/2} \langle \mathbf{xv}^T\rangle \mathcal{A}^{1/2} =  \mathcal{A}^{1/2} {C}_{xv} \mathcal{A}^{1/2}$, and the commutator transforms as $[\mathcal{A}, \nabla_{\mathbf{X}} \nabla_{\mathbf{X}} \tilde{U}]=\mathcal{A}^{-1/2}[\mathcal{A}, \nabla_{\mathbf{x}} \nabla_\mathbf{x} U]\mathcal{A}^{-1/2}=\frac{1}{\tilde{T}}\mathcal{A}^{1/2}[\nabla_\mathbf{x}\nabla_\mathbf{x}U, \mathcal{T}]\mathcal{A}^{1/2}$. The force field in this case is $\mathbf{F}=-\nabla_{\mathbf{x}}\nabla_{\mathbf{x}}U\cdot\mathbf{x}$ and we denote $\Phi\equiv\nabla_{\mathbf{x}}\nabla_{\mathbf{x}}U$.
\eqref{eq:covariance_source} and \eqref{eq:OU_EPR_result} then read
\begin{equation}
\begin{split}
        2\xi^{2} &\left(\Phi\,C_{xv} + C_{xv}\,\Phi\right)\\ &+ m\left(\Phi^{2}C_{xv}
    - 2\,\Phi\,C_{xv}\,\Phi
    + C_{xv}\Phi^{2}\right)
    = 2k_{B}\xi\left[\Phi,\,\mathcal{T}\right],
\end{split}
\end{equation}
and
\begin{equation}
        \langle\dot{S}\rangle = -\frac{1}{2}\,\mathrm{Tr}\left(\left[\mathcal{T}^{-1},\,\Phi\right]C_{xv}\right).
\end{equation}
While less transparent than the mechanical interpretation where the curl of the force field acts as a driving source, this description highlights the role of the coupling between DoFs with different temperatures in driving the system out of equilibrium.

\subsection*{Stall Force} \label{sec:equilibration_by_forcing}
As a final application, we show how the curl-force description of a multi-reservoir system can be used to easily determine the (non-conservative) force required to equilibrate it. This part is inspired by Refs.~\cite{ivlev2015statistical, loos2020irreversibility}, where it was shown that systems with specific non-reciprocal interactions that are coupled to multiple heat baths can in fact be in thermal equilibrium. Here, we generalize this statement, and explicitly construct the required non-conservative force and equilibrium distribution. As we show below, the solution is \emph{not} unique, and given $\mathcal{T}$ and $U(\mathbf{x})$, there is some freedom in choosing the target equilibrium distribution (and the corresponding force field).

\subsubsection*{Equilibrating curl-force}
We now explicitly formulate the problem. Given a non-equilibrium system where a particle with a non-separable potential $U(\mathbf{x})$ is coupled to multiple reservoirs  (such as the two examples studied in the applications section), can a non-conservative force be applied on the system in a manner that would exactly cancel any circulation, relaxing the system to equilibrium?

To answer this question, we consider a multi-reservoir system subject to curl-force $-\mathcal{B}\nabla_{\mathbf{x}}U$, where $\mathcal{B}$ is a symmetric invertible matrix that is to be determined. The Klein-Kramers equation describing the system's evolution in this case is
\begin{equation}\label{eq:klein_kramers_multibath_with_curl}
\begin{split}
  \frac{\partial f_t(\mathbf{x}, \mathbf{v})}{\partial t} =  &\frac{\mathcal{B}\nabla_{\mathbf{x}} U\cdot \nabla_{\mathbf{v}}f_t}{m}-\nabla_{\mathbf{x}} f_t \cdot \mathbf{v} \\ &+ \frac{\xi}{m} \nabla_{\mathbf v} \cdot\left(\mathbf{v} f_t + \frac{k_{\rm B}}{m} \mathcal{T}\cdot\nabla_{\mathbf{v}}f_t\right),
  \end{split}
\end{equation}
and we can still apply the transformation of \eqref{eq:canonical_transformation}. The transformed system becomes
\begin{equation}\label{eq:curl_curl_description}
\begin{split}
      \frac{\partial \tilde{f}_t(\mathbf{X}, \mathbf{V})}{\partial t} = &\frac{\tilde T (\mathcal{T}^{-1/2}\mathcal{B}\mathcal{T}^{-1/2}\nabla_{\mathbf{X}} \tilde{U}) \cdot \nabla_{\mathbf{V}}\tilde{f}_t}{m}-\nabla_{\mathbf{X}} \tilde{f}_t \cdot \mathbf{V} \\ &\qquad+ \frac{\xi}{m} \nabla_{\mathbf V} \cdot\left(\mathbf{V} \tilde{f}_t + \frac{k_{\rm B} \tilde T}{m} \nabla_{\mathbf{V}}\tilde{f}_t\right).
\end{split}
\end{equation}
In this case, the matrix $\mathcal{A}=\tilde T\mathcal{T}^{-1/2}\mathcal{B}\mathcal{T}^{-1/2}$ takes the role of the anisotropy matrix in the transformed, single-bath system. A sufficient condition for \eqref{eq:curl_curl_description} to describe an equilibrium system is $\mathcal{A}=\mathbb{I}$\footnote{More generally, we can choose any $\mathcal{A}$ satisfying $[\mathcal{A}, \nabla \nabla \tilde{U}]=0$. However, for a constant $\mathcal{A}$ this condition is quite limiting (at least beyond the case of an OU process where the Hessian is constant matrix). Overall, we expect $\mathcal{A}\propto\mathbb{I}$ to be the most general solution unless the Hessian has a specific structure that can be exploited.}, so 
\begin{equation}
    \mathcal{B} = \tilde{T}^{-1} \mathcal{T}.
    \label{the_B}
\end{equation}
The curl-force in the single-bath description reduces to a gradient force $\nabla_{\mathbf{X}}\tilde{U}$ and the equilibrium solution is given by the Boltzmann distribution 
\begin{equation}
    \tilde{f}_{\rm eq}(\mathbf{X}, \mathbf{V}) = \frac{1}{\tilde{Z}}\exp\left(-\frac{\frac{1}{2}m\mathbf{V}^2+\tilde{U}(\mathbf{X})}{k_{\rm B} \tilde{T}}\right),
\end{equation}
where $\tilde{Z}$ is the partition function which is the corresponding normalization.

Finally, we can transform $\tilde{f}_{\rm eq}$ back to the multi-reservoir coordinates to obtain the stationary solution of \eqref{eq:klein_kramers_multibath_with_curl} as
\begin{equation}\label{eq:multi_bath_equilibrium}
    f_{{\rm eq},\tilde{T}}(\mathbf{x}, \mathbf{v}) = \frac{1}{Z}\exp\left(-\frac{m}{2k_{\rm B}}\mathbf{v}\mathcal{T}^{-1}\mathbf{v}-\frac{U(\mathbf{x})}{k_{\rm B} \tilde{T}}\right),
\end{equation}
with $Z$ the corresponding normalization. The first term in the exponential is the multi-reservoir Maxwell distribution, such that the velocity distribution corresponds to that of a system where each DoF equilibrates separately with its own reservoir. The spatial part, however, corresponds to the Boltzmann distribution for a system in contact with a \emph{single} effective temperature $\tilde{T}$. 

We note that \eqref{eq:multi_bath_equilibrium} is a genuine equilibrium solution: not only is it a steady state solution of corresponding Klein-Kramers equation, but is also characterized by zero entropy production rate and vanishing heat and work in any DoF. However, the solution is \emph{not} unique, i.e., {any choice of $\tilde{T}$ in the definition of the curl force $-\mathcal{B}\nabla_{\mathbf{x}}U$ with $\mathcal{B}$ in \eqref{the_B} is valid, and a particular choice determines the temperature $\tilde{T}$ to which the spatial DoF $\mathbf{x}$ equilibrate to.}

\subsubsection*{Examples}
The above method can be directly applied to both gyrator systems studied above. For the harmonic Brownian gyrator, the total non-conservative force\footnote{This includes the gradient force $-\nabla U$ naturally applied on the system by the potential. Subtract $-\nabla U$ to obtain the magnitude of the additional external force that needs to be applied in order for the system to equilibrate.} required to oppose the circulation is
\begin{equation}
    \mathbf{F}_{\rm eq}^\text{harmonic} = -\frac{1}{\tilde{T}}\begin{pmatrix}
        T_x (kx+uy) \\
        T_y (ky+ux)
    \end{pmatrix},
\end{equation}
whereas for the {azimuthal-symmetric} gyrator
\begin{equation}
    \mathbf{F}_{\rm eq}^\text{azimuthal} = -\frac{k}{\tilde{T}} \left(1 - \frac{r_0}{\sqrt{x^2+y^2}}\right)\begin{pmatrix}
        T_x x \\
        T_y y
    \end{pmatrix}.
\end{equation}
In both cases, $\nabla_{\mathbf{x}}\times\mathbf{F}_{\rm eq}$ is in the direction opposite to $\nabla_{\mathbf{X}}\times\tilde{\mathbf{F}}$ of the gyrating system in the curl-force description, as expected (the force \emph{opposes} the gyration naturally occurring in the original non-equilibrium system). As demonstrated through the second example, the method is equally applicable to linear and non-linear force fields.

\section*{Discussion and outlook}\label{sec:discussion}
We studied the underdamped nonequilibrium dynamics of 
a system 
coupled to heat reservoirs with different temperatures and introduced a canonical transformation mapping it onto a system coupled to a \emph{single} reservoir with an augmented force. The nonequilibrium nature of the transformed system persists due to an emergent deterministic curl-force, which is non-conservative. The transformed dynamics is generated by a Hamiltonian with an anisotropic kinetic energy term \cite{berry2015hamiltonian}, with anisotropy $\mathcal{A}\propto\mathcal{T}^{-1}$ fixed by the temperatures of the original reservoirs.
The mapping unifies two distinct mechanisms of driving a system out of equilibrium---coupling to several reservoirs, and applying a non-conservative force---and provides a mechanical interpretation of the rotational motion typically observed in multi-temperature systems. The energetics of the two descriptions are fundamentally different: In the multi-reservoir picture heat flows from the hot to the cold reservoir and no work is performed; in the curl-force picture positive work is always applied on the system and is fully dissipated into the single bath, as required by the second law. {Meanwhile, the entropy production is consistently equal in both representations.}

We stressed that the Hamiltonian structure of the transformed system is not sufficient for equilibration; the system does \emph{not} relax to the Boltzmann-Gibbs distribution $\tilde{f}_{\rm eq}\propto e^{-\tilde{H}/k_{\rm B}\tilde{T}}$ associated with $\tilde{H}$ and the single bath temperature $\tilde{T}$, 
{because of a non-conservative force emerging in the transformation while interactions with the bath remain standard.}
{The transformed description is nevertheless useful as it directly provides qualitative insight about the multi-reservoir system, in particular about the nonequilibrium stationary circulation.  Specifically 
for the \emph{Brownian gyrator} and the \emph{azimuthal-symmetric gyrator}, the circulation direction is encoded in the curl force sign, namely the sign of $\nabla\times\tilde{\mathbf{F}}$, which does \emph{not} require any knowledge about the steady-state distribution}. The transformed curl-force approach has the benefit of being intuitive and easier to grasp in comparison to alternative ways of qualitatively explaining non-equilibrium phenomena, such as through an analysis of the angular dependence of the diffusion tensor \cite{abdoli2025quadrupolar}. It is clear, however, that $\nabla\times\tilde{\mathbf{F}}$ cannot {fully (and in particular quantitatively)} describe the currents in the thermal system as it lacks information regarding the steady-state distribution. Note, however, that similar quantities proposed for the characterization of non-trivial dynamics arising in such classes of systems, such as the curl of the probability current \cite{dotsenko2013two}, are not necessarily more informative, as they tend to be involved and often lack a clear physical significance. Hence $\nabla \times \tilde{\mathbf{F}}$, which is easily computed for arbitrarily complex systems, may sufficiently describe certain desired properties of a nonequilibrium system. We further note that if $\nabla \times \tilde{\mathbf{F}} = 0$ in the curl-force representation, then the system is in thermal equilibrium with the bath. In the multiple-bath description, this is reflected in each DoF independently equilibrating with its own bath.

For the exactly solvable $N$-dimensional Ornstein--Uhlenbeck process, the qualitative role of $\nabla \times \tilde{\mathbf{F}}$ in driving the system out of equilibrium was made precise. The steady-state covariance matrix was obtained in closed form in terms of the equilibrium result corrected by the generalized curl $[\mathcal{A},\nabla\nabla\tilde{U}]$, and a similar result was obtained for the EPR, highlighting that $[\mathcal{A},\nabla\nabla\tilde{U}]=0$ directly implies a vanishing entropy production {(and hence equilibrium)}. While this fact is trivial, it is not immediately apparent from the general form of the EPR in \eqref{eq:EPR} and \eqref{eq:EPR_transformed}. {Back-transformed to the multi-reservoir description, these exact results highlight the role of the coupling between the reservoirs in driving the system out of thermal equilibrium.}

{We have furthermore addressed the ``stall force'' phenomenon, whereby a system coupled to multiple reservoirs equilibrates upon the application of non-conservative forces. We provided a general expression for the required force field and derived the corresponding equilibrium distribution.}

Next, we highlight two related works and comment on the conceptual and practical similarities and differences. First, it is noteworthy that a transformation similar to \eqref{eq:canonical_transformation} was introduced by Loos and Klapp \cite{loos2020irreversibility} in the context of linear non-reciprocal systems coupled to multiple heat baths. 
A second idea conceptually similar to our curl-force description is the \emph{Landscape-Flux Framework} for nonequilibrium systems \cite{wang2008potential} and in particular its application to systems coupled to multiple heat baths \cite{wu2021landscape}. Within this framework, the system's stochastic dynamics is decomposed into an effective generalized potential derived from the steady-state probability density 
(generally, this potential can be a function of all phase space variables $\mathbf{x},\mathbf{v}$ and is not restricted to having only a spatial dependence), accompanied by a curl flux which can be seen as the result of an effective non-conservative force \cite{wang2008potential}. The main conceptual difference to our work is that the Landscape-Flux Framework is constructed \emph{a posteriori}, in the sense that the definitions of the effective potential and curl flux \emph{are based on} the NESS distribution itself, which is a practical limitation since exact solutions are typically unknown. Our framework evokes the non-conservative force description from first principles, without having to first solve the stochastic dynamics. This allows us to apply the transformation to systems with arbitrary potentials $U(\mathbf{x})$, and as we demonstrate in one of the examples studied in the applications section, we can qualitatively describe some properties of the stochastic system even in the case where an exact solution is unknown. We further note that the landscape-flux framework for systems with multiple reservoirs \cite{wu2021landscape} does not eliminate the coupling to the multiple reservoirs, but rather highlights the role of the multiple temperatures in generating fluxes and modifying the steady-state distribution.

We now discuss applications and provide an outlook for future research. One especially promising application of our transformation is in experimental realizations of systems with multiple heat baths. A genuine multi-temperature Brownian
particle does not exist, since simultaneous contact with reservoirs at multiple temperatures is difficult to give a microscopic meaning for a single particle. While this has been overcome in experiments by considering analog systems \cite{chiang2017electrical, ciliberto2013heat, argun2017experimental, berut2014energy, lin2022stochastic, berut2016stationary, martins2026role}, we note that curl-forces can be generated with existing setups such as optical tweezers \cite{roichman2008influence,roichman2008optical, sun2009brownian, wu2009direct, liberal2013near, sukhov2017non, PhysRevLett.122.183901, PhysRevLett.127.038003} and feedback traps \cite{cohen2005control, lopez2008realization, toyabe2010experimental, jun2012virtual}, and thus the transformed Brownian motion process can be implemented in the laboratory in a genuine thermal setup. In the context of our results, 
the multi-temperature experiment could therefore be emulated by a single-bath and curl-force.

Several directions remain open. Notably, the Brownian gyrator can be mapped to an information engine \cite{leighton2024information}. In the context of our work, it is worthwhile considering how the curl-force description of the system relates to the feedback-control one, and more generally how these two methods of driving a system out of equilibrium relate to one another \cite{PhysRevResearch.5.023080}. It would also be interesting to apply our transformation to gyrators with other potentials \cite{chang2021autonomous}, which could provide general insights beyond those obtained for the harmonic and azimuthal-symmetric potentials studied here. It would furthermore be interesting to study other decompositions of nonequilibrium processes in the context of the curl-force setting, for example in the context of iso-dissipation forces \cite{Duong_2023}, canonical conservative dynamics \cite{Qian}, as well as bivectorial \cite{graham1977covariant,Yang_2021} and circulation \cite{Min_1999} decompositions.
{More broadly, it would be interesting to 
explore systems with general Hamiltonian curl forces. For example, the identification of conserved quantities \cite{yavari2025darboux} could be a step towards a more general theory of the statistical mechanics of such systems. Further studying other types of ``peculiar Hamiltonians'' and how they relate to different classes of known nonequilibrium systems could eventually lead to a more unified theory for nonequilibrium statistical mechanics.}
Finally, it is intriguing to construct a kinetic theory for a closed system consisting of particles with a non-isotropic kinetic energy term, namely when the bath itself has a curl-force Hamiltonian. Could such a system relax strongly (i.e.\ is the corresponding Markov semigroup a contraction), and if so, would it relax to the corresponding equilibrium or a nonequilibrium steady state?

\acknowledgments
{We acknowledge financial support European Research
Council (ERC) under the European Union’s Horizon Europe
research and innovation program (Grant Agreement
No. 101086182 to A. G.), the German Research
Foundation (DFG) through the Heisenberg Program (Grant
No. GO 2762/4-1 to A. G.), the ISF, grant no. 232/23 (to O.R.), the Minerva foundation (to O.R.), and the Ministry of Research and Culture in Lower Saxony and the Volkswagen Foundation within the
program ”Research Cooperations Lower Saxony–Israel”
(jointly to A.G. and O.R.).
We thank Roi Holtzman and Shahaf Aharony Shapira for helpful discussions.}

\bibliography{main}

\end{document}


\title{SI Appendix: Hamiltonian curl-forces in systems coupled to multiple thermal reservoirs}

\author{Omer Chor}
\affiliation{Department of Physics of Complex Systems, Weizmann Institute of Science, Rehovot, 761001, Israel}
\author{Alja\v{z} Godec}
\affiliation{Mathematical Physics and Stochastic Dynamics, Institute of Physics,
University of Freiburg, 79104 Freiburg im Breisgau, Germany}
\affiliation{Mathematical bioPhysics Group, Max Planck Institute for Multidisciplinary Sciences, G\"ottingen 37077, Germany}
\author{Oren Raz}
\affiliation{Department of Physics of Complex Systems, Weizmann Institute of Science, Rehovot, 761001, Israel}

\date{\today}

\maketitle

\section{Perturbative solution of the Klein-Kramers equation with anisotropic temperatures in the strongly-damped limit}\label{sec:overdamped_expansion}

Starting from the Klein-Kramers equation (Eq.~4 in the main text),  we are looking for a perturbative steady-state solution $f_{\rm s}(\mathbf{x}, \mathbf{v}) = f_0+\varepsilon f_1+ O(\varepsilon^2)$ in the high-friction (overdamped) limit $0<\frac{m}{\xi}\ll1$, where we denote the small parameter $\varepsilon=\frac{m}{\xi}$. The steady-state equation is
 \begin{align}
    0 &= \varepsilon\left(\frac{1}{m}\nabla_{\mathbf{x}} {U} \cdot \nabla_{\mathbf{v}}f_{\rm s} -\nabla_{\mathbf{x}} f_{\rm s} \cdot {\mathbf{v}}\right) +  \nabla_{\mathbf v}\cdot(\mathbf{v} f_{\rm s} + \frac{k_B }{m} \mathcal{T}\cdot\nabla_{\mathbf{v}}f_{\rm s}) \\
    &=\varepsilon \{{H}, f_{\rm s}\} + \mathcal{L}f_{\rm s}, \nonumber 
 \end{align}
 where $\mathcal{L}f_{\rm s} = \left[\nabla_{\mathbf v}\cdot(\mathbf{v} + \frac{k_B}{m} \mathcal{T}\cdot\nabla_{\mathbf{v}})\right]f_{\rm s}$.

 \subsubsection*{$f_0$ and formulation of the eigenvalue problem} Since coupling between the degrees of freedom is induced by the Liouville flow, to this order the solution is simply the multi-temperature Maxwell distribution $g_0(\mathbf{v})$, multiplied by an arbitrary function of coordinates $\rho(\mathbf{x})$,
 \begin{equation}
     f_0(\mathbf{x},\mathbf{v}) = \rho(\mathbf{x})g_0(\mathbf{v}) = \rho(\mathbf{x})\left(\frac{m}{2\pi k_B}\right)^{N/2} |\mathcal{T}|^{-1/2} \exp\left( - \frac{m\mathbf{v}^T\mathcal{T}^{-1}\mathbf{v}}{2k_B}\right).
 \end{equation}
It is then useful to decompose the full solution as
\begin{equation}
    f_s(\mathbf{x}, \mathbf{v}) = \phi(\mathbf{x}, \mathbf{v})g_0(\mathbf{v}),
\end{equation}
because
 \begin{equation}
     \mathcal{L}\left(\phi g_0\right) = \left[\frac{k_B}{m}\nabla_{\mathbf{v}}\cdot\left(\mathcal{T}\nabla_{\mathbf{v}} \phi\right) - \mathbf{v}\cdot\nabla_{\mathbf{v}} \phi\right] g_0(\mathbf{v}). \label{eq:LV_action}
 \end{equation}
 Rescaling $\mathbf{v}$ as
 \begin{equation}
      \mathbf{w}  \equiv \left(\frac{m}{k_B}\mathcal{T}^{-1}\right)^{1/2}\mathbf{v},
 \end{equation}
 we obtain
 \begin{equation}
     \mathcal{L}(\phi g_0) = (\nabla^2_{\mathbf{w}}\phi - \mathbf{w}\cdot\nabla_{\mathbf{w}}\phi)g_0.
 \end{equation}
 Therefore, we identify the eigenvalue problem $\mathcal{L}(\phi g_0) = -n \phi g_0$ ($n=0,1,2,\ldots$) as the Hermite differential equation multiplied by $g_0$, such that Hermite polynomials are its eigenfunctions.

\subsubsection*{Solvability condition}
 Since $\mathcal{L}f$ is a total divergence in $\mathbf{v}$, integrating the order-$\varepsilon^n$ equation $\mathcal{L}f_n = -\{{H}, f_{n-1}\}$ over velocities yields the solvability (Fredholm) condition
 \begin{equation}
     \int \{{H}, f_{n-1}\}\, d\mathbf{v} = 0 \quad \text{pointwise in } \mathbf{x}. \label{eq:solvability_general}
 \end{equation}
 
 \subsubsection*{Solving for ${f_1}$} 
 The equation for order $\varepsilon^1$ is
 \begin{equation}
     \mathcal{L}f_1 = -\{{H}, f_0\} = \left(\frac{\mathcal{T}^{-1}\nabla_{\mathbf{x}} {U}}{k_B}\rho(\mathbf{x}) + \nabla_\mathbf{x}\rho\right)\cdot\mathbf{v} g_0.
 \end{equation}
The solvability condition is therefore satisfied for any $\rho(\mathbf{x})$ as this is an odd function of $\mathbf{v}$.
Since the first Hermite polynomial is $\mathrm{He}_1(\mathbf{v})=\mathbf{v}$, the Hermite equation is $\mathcal{L}(\mathbf{v}g_0) = -\mathbf{v}g_0$, and we immediately identify the solution as 
\begin{equation}
    f_1(\mathbf{x}, \mathbf{v}) = - \left(\frac{\mathcal{T}^{-1}\nabla_{\mathbf{x}} {U}}{k_B}\rho(\mathbf{x}) + \nabla_\mathbf{x}\rho(\mathbf{x})\right)\cdot\mathbf{v} g_0 \equiv \bm{\alpha}(\mathbf{v})\cdot\mathbf{v}g_0.
\end{equation}
We note that the stationary spatial probability flux is
\begin{equation}
    \mathbf{J}(\mathbf{x}) = \int \, \mathbf{v} f_s d\mathbf{v} = \varepsilon \int \, \mathbf{v} (\bm{\alpha}(\mathbf{x})\cdot\mathbf{v}) g_0 d\mathbf{v} + O(\varepsilon^2) = \varepsilon \frac{k_B}{m}\mathcal{T} \bm{\alpha}(\mathbf{x}) + O(\varepsilon^2).
\end{equation}
Integrating the Klein-Kramers equation over $\mathbf{v}$, the stationarity condition is
\begin{equation}
    \nabla_\mathbf{x} \cdot \mathbf{J}(\mathbf{x}) = \nabla_{\mathbf{x}}\cdot\left(\frac{\nabla_{\mathbf{x}} {U}}{\xi}\rho(\mathbf{x}) + \frac{k_B \mathcal{T}}{\xi}\nabla_\mathbf{x}\rho(\mathbf{x})\right)= 0,
\end{equation}
which is the stationary Smoluchowski equation for a system with anisotropic diffusion matrix $k_B\mathcal{T}/\xi$, so the multiple-reservoir description carries on to the strongly-damped limit.

\section{Steady-state solution of the N-dimensional OU process and the role of the (generalized) curl of the force field}
\label{sec:N_dim}

In this Appendix we solve for the steady state of the curl-force Klein-Kramers
dynamics in closed form for a quadratic potential in $N$ dimensions, and obtain explicit expressions for all blocks of the covariance matrix $\tilde{C}$. While these can be obtained directly by using computer algebra \cite{gowda2021high} to solve the Lyapunov equation (see below), and are easily computed numerically for given system parameters, it is instructive to explicitly derive the solution, because the expressions, even for the two dimensional case, are typically quite intricate. We further note that computer algebra solutions beyond two dimensions (four dimensional phase space) are generally quite slow, as the generated expressions tend to be quite complicated. Here we express the deviation of the steady-state covariance matrix from the equilibrium solution in a manner that highlights the role of the asymmetric part of the force field.
The two-dimensional case of the Brownian gyrator treated in the main text is recovered at the end as a special case.

\subsection*{Setup and block structure of the Lyapunov equation}

We consider an underdamped particle with mass $m$ in $N$ dimensions subject to the curl force
$\tilde{\mathbf{F}} \equiv -\tilde{\Phi} \mathbf{X} = -\mathcal{A}\tilde{\mathcal{U}}\,\mathbf{X}$, with $\mathcal{A}$ the symmetric, positive-definite anisotropy matrix and $\tilde{\mathcal{U}}\equiv\nabla\nabla \tilde{U}$ the Hessian of the potential $\tilde{U}(\mathbf{X}) = \tfrac{1}{2}\mathbf{X}^{T}\tilde{\mathcal{U}}\mathbf{X}$, coupled to a single bath at temperature $\tilde{T}$ with drag coefficient $\xi$. In the phase-space coordinates $\mathbf{Z} = (\mathbf{X},\mathbf{V})^{T}$ the dynamics is the Ornstein-Uhlenbeck process
\begin{equation}
    d\mathbf{Z} = -\mathcal{F}\,\mathbf{Z}dt + \sqrt{2\mathcal{D}}d\mathbf{W}(t),
    \qquad
    \left\langle dW_i(t)\,dW_j(t')\right\rangle
    = \delta_{ij}\,\delta(t-t')dtdt',
    \label{eq:appN-ou}
\end{equation}
with the drift and diffusion matrices
\begin{equation}
    \mathcal{F} = \begin{pmatrix}
        0 & -\mathbb{I} \\[2pt]
        \tilde{\Phi}/m & (\xi/m)\,\mathbb{I}
    \end{pmatrix},
    \qquad
    \mathcal{D} = \begin{pmatrix}
        0 & 0 \\[2pt]
        0 & \dfrac{\xi k_{B}\tilde{T}}{m^{2}}\,\mathbb{I}
    \end{pmatrix}
\end{equation}
where $0$ and $\mathbb{I}$ are the $N\times N$ zero and unit matrices, respectively. It is noteworthy that force matrix $\tilde{\Phi}$ is the product of two symmetric matrices and is therefore \emph{not} symmetric in general, its antisymmetric part being half the commutator, $\tilde{\Phi} - \tilde{\Phi}^{T} = [\mathcal{A},\tilde{\mathcal{U}}]$.
The stationary covariance matrix (which exists if the deterministic dynamics, $m\ddot{\mathbf{X}} \equiv -\tilde{\Phi} \mathbf{X}$, is stable \cite{risken1996fokker})
\begin{equation}
    \tilde{C} = \begin{pmatrix}
        \tilde{C}_{XX} & \tilde{C}_{XV} \\
        \tilde{C}_{XV}^T & \tilde{C}_{VV}
    \end{pmatrix}
\end{equation}
is a $2N\times2N$ symmetric matrix, written here in terms of its $N\times N$ blocks. It obeys the Lyapunov equation $\mathcal{F}\tilde{C} + \tilde{C}\mathcal{F}^{T} = 2\mathcal{D}$, given explicitly by
\begin{equation}
    \begin{pmatrix}
        -\big(\tilde{C}_{XV} + \tilde{C}_{XV}^{T}\big)
        & \dfrac{1}{m}\,\tilde{C}_{XX}\tilde{\Phi}^{T} + \dfrac{\xi}{m}\,\tilde{C}_{XV} - \tilde{C}_{VV}
        \\[10pt]
        \dfrac{1}{m}\,\tilde{\Phi} \tilde{C}_{XX} + \dfrac{\xi}{m}\,\tilde{C}_{XV}^{T} - \tilde{C}_{VV}
        & \dfrac{1}{m}\big(\tilde{\Phi} \tilde{C}_{XV} + \tilde{C}_{XV}^{T}\tilde{\Phi}^{T}\big)
        + \dfrac{2\xi}{m}\,\tilde{C}_{VV}
    \end{pmatrix}
    =
    \begin{pmatrix}
        0 & 0 \\[6pt]
        0 & \dfrac{2\xi k_{B}\tilde{T}}{m^{2}}\,\mathbb{I}
    \end{pmatrix}.
    \label{eq:appN-lyapunov-blocks}
\end{equation}
From symmetry of the covariance matrix $\tilde{C}$, the blocks $\tilde{C}_{XX}, \tilde{C}_{VV}$ are symmetric, and the equation has three independent block conditions.

First, the position-position block $\tilde{C}_{XV} + \tilde{C}_{XV}^{T} = 0$ implies the $\tilde{C}_{XV,ij} = \langle X_{i}V_{j}\rangle = \tfrac{1}{2}\langle X_{i}V_{j} - X_{j}V_{i}\rangle $ is antisymmetric. $\tilde{C}_{XV}$ collects the $N(N-1)/2$ independent planar circulations of the steady state. Specifically, in two dimensions the circulation is described by a single number $\omega = \langle XV_Y - YV_X\rangle$, with
\begin{equation*}
    \tilde{C}_{XV}^{(2d)} = \frac{\omega}{2}\begin{pmatrix}
        0 & 1 \\
        -1 & 0
    \end{pmatrix}.
\end{equation*}
In terms of $\tilde{C}_{XV}$, the velocity-velocity block reads
\begin{equation}
    \tilde{C}_{VV} = \frac{k_{B}\tilde{T}}{m}\,\mathbb{I}
    - \frac{1}{2\xi}\left(\tilde{\Phi} \tilde{C}_{XV} - \tilde{C}_{XV}\tilde{\Phi}^{T}\right),
    \label{eq:appN-Cv}
\end{equation}
i.e., the velocity fluctuations are given by the equilibrium (Maxwell-Boltzmann) value with a correction due to the circulation $\tilde{C}_{XV}$.
Finally, the position-velocity block is given by
\begin{equation}
    \tilde{C}_{XX}\tilde{\Phi}^T = m\,\tilde{C}_{VV} - \xi \tilde{C}_{XV}.
    \label{eq:appN-Cx}
\end{equation}
Hence, the steady state for a given $\tilde{\Phi}$ is fully determined by the $N(N-1)/2$ independent circulations $\tilde{C}_{XV,ij}$. 

To proceed, we first eliminate $\tilde{C}_{XX}$ by using the symmetry of $\tilde{C}_{XX}, \tilde{C}_{VV}$ and the antisymmetry of $\tilde{C}_{XV}$, obtaining
\begin{equation}
    m\left(\tilde{\Phi} \tilde{C}_{VV} - \tilde{C}_{VV}\tilde{\Phi}^{T}\right)
    = \xi\left(\tilde{\Phi} \tilde{C}_{XV} + \tilde{C}_{XV}\tilde{\Phi}^{T}\right).
    \label{eq:appN-closure}
\end{equation}
 It remains to express the left-hand side through $\tilde{C}_{XV}$ using \eqref{eq:appN-Cv}. Multiplying \eqref{eq:appN-Cv} by
$\tilde{\Phi}$ from the left and, separately, by $\tilde{\Phi}^{T}$ from the right, and subtracting, we obtain
\begin{align}
    \tilde{\Phi} \tilde{C}_{VV} - \tilde{C}_{VV}\tilde{\Phi}^{T}
    &= \frac{k_{B}\tilde{T}}{m}\big(\tilde{\Phi} - \tilde{\Phi}^{T}\big)
    - \frac{1}{2\xi}\left(\tilde{\Phi}^{2}\tilde{C}_{XV} - 2\,\tilde{\Phi} \tilde{C}_{XV}\tilde{\Phi}^{T}
    + \tilde{C}_{XV}(\tilde{\Phi}^{T})^{2}\right) \nonumber \\
    &= \frac{k_{B}\tilde{T}}{m}\,[\mathcal{A},\tilde{\mathcal{U}}]
    - \frac{1}{2\xi}\left(\tilde{\Phi}^{2}\tilde{C}_{XV} - 2\,\tilde{\Phi} \tilde{C}_{XV}\tilde{\Phi}^{T}
    + \tilde{C}_{XV}(\tilde{\Phi}^{T})^{2}\right).
\end{align}
Substituting into
\eqref{eq:appN-closure},
\begin{equation}
    k_{B}\tilde{T}\,[\mathcal{A},\tilde{\mathcal{U}}]
    - \frac{m}{2\xi}\left(\tilde{\Phi}^{2}\tilde{C}_{XV} - 2\,\tilde{\Phi} \tilde{C}_{XV}\tilde{\Phi}^{T}
    + \tilde{C}_{XV}(\tilde{\Phi}^{T})^{2}\right)
    = \xi\left(\tilde{\Phi} \tilde{C}_{XV} + \tilde{C}_{XV}\tilde{\Phi}^{T}\right).
\end{equation}
Rearranging, we arrive at an equation for $\tilde{C}_{XV}$ in terms of the force $\tilde{\Phi}=\mathcal{A}\tilde{\mathcal{U}}$ and its generalized curl $[\mathcal{A}, \tilde{\mathcal{U}}]$,
\begin{align}
    2\xi^{2}\left(\tilde{\Phi} \tilde{C}_{XV} + \tilde{C}_{XV}\tilde{\Phi}^{T}\right)
    + m\left(\tilde{\Phi}^{2}\tilde{C}_{XV} - 2\tilde{\Phi} \tilde{C}_{XV}\tilde{\Phi}^{T} + \tilde{C}_{XV}(\tilde{\Phi}^{T})^{2}\right)
    & = 2 k_B \tilde{T} \xi (\tilde{\Phi}-\tilde{\Phi}^T)\nonumber \\&= 2k_{B}\tilde{T}\xi[\mathcal{A},\tilde{\mathcal{U}}].
    \label{eq:appN-master}
\end{align}
Both sides are antisymmetric, so this is a linear system of $N(N-1)/2$
equations for the $N(N-1)/2$ independent entries of $\tilde{C}_{XV}$, sourced by the commutator: if $[\mathcal{A},\tilde{\mathcal{U}}] = 0$ (equivalently, if $\tilde{\Phi}=\tilde{\Phi}^T$) then $\tilde{C}_{XV} = 0$ and the steady state is the equilibrium one, while any nonzero commutator induces circulation (recall that the commutator of $\mathcal{A}$ and $\tilde{\mathcal{U}}\equiv\nabla\nabla \tilde{U}$ is the N-dimensional generalization of the curl of the force field).

We now proceed to solving \eqref{eq:appN-master} by a spectral decomposition of $\tilde{\Phi}$ \cite{risken1996fokker}. We note that the underdamped dynamics solved here are a special (and simpler) case of the general solution of the OU process solved by Risken \cite{risken1996fokker} because of the specific block form that $\mathcal{F}$ takes in this case.

\subsection*{Solution by spectral decomposition}
\eqref{eq:appN-master} is solved by passing to the normal modes of
the deterministic dynamics. The noiseless, frictionless equation of motion
$m\ddot{\mathbf{X}} = -\tilde{\Phi}\mathbf{X}$ is solved by the eigenvalue problem $\tilde{\Phi}\mathbf{r}_{i} = \lambda_{i}\mathbf{r}_{i}$. We note that while $\tilde{\Phi}$ is not symmetric, it is similar to the (symmetric) Hessian $\mathcal{U}$ of the same OU process in the multi-bath description (after transforming the potential according to the canonical transformation Eq.~9 in the main text), such that it has a real spectrum and complete eigenbasis. We further note again that a steady-state solution exists only if the deterministic dynamics are stable, that is the eigenvalues must also be positive (equivalently, $\mathcal{U}$ is positive definite).
We expand the solution in terms of $\tilde{\Phi}$'s biorthogonal set of right $\mathbf{r}_{i} = \mathcal{A}^{1/2} \mathbf{e}_{i}$ and left $\bm{\ell}_{i} = \mathcal{A}^{-1/2}\mathbf{e}_{i}$ eigenvectors, where ${\mathbf{e}_i}$ is the orthonormal eigenbasis of $\mathcal{U}$, such that $\bm{\ell}_{i}\cdot\mathbf{r}_{j} = \delta_{ij}$. We then define the matrix $R=(\mathbf{r}_1,\mathbf{r}_2, \ldots,\mathbf{r}_N)$ whose columns are the right eigenvectors, and the diagonal matrix $\Lambda=\mathrm{diag}(\lambda_1, \lambda_2,\ldots,\lambda_N)$.

Expanding any matrix as $M = R\hat{M}R^{T}$, it is useful to transform the non-trivial block of the diffusion matrix as
\begin{equation}
    \hat{\mathcal{D}}_{VV} = R^{-1} D_{VV} (R^T)^{-1} = \frac{\xi k_B \tilde{T}}{m^2}(R^T R)^{-1} = \frac{\xi k_B \tilde{T}}{m^2}G, \qquad G \equiv \left(R^{T}R\right)^{-1}.
\end{equation}
One also finds
\begin{equation}
    \tilde{\Phi} M + M\tilde{\Phi}^{T} = \sum_{i,j}(\lambda_i+\lambda_j)\hat{M}_{ij}\mathbf{r}_i\mathbf{r}_j^T,
    \qquad
    \tilde{\Phi}^{2}M - 2\tilde{\Phi} M\tilde{\Phi}^{T} + M(\tilde{\Phi}^{T})^{2}
    = \sum_{i,j}(\lambda_i-\lambda_j)^2 \hat{M}_{ij}\mathbf{r}_i\mathbf{r}_j^T,
\end{equation}
and noting that $\tilde{\Phi}=R\Lambda R^{-1}=R\Lambda GR^T$ (the last step follows from the definition of $G$), the source term in \eqref{eq:appN-master} becomes
\begin{equation}
    [\mathcal{A},\tilde{\mathcal{U}}] = \tilde{\Phi} - \tilde{\Phi}^{T} = R[\Lambda,G]R^{T},
    \qquad
    \left([\Lambda,G]\right)_{ij} = (\lambda_{i}-\lambda_{j})\,G_{ij}.
    \label{eq:appN-source}
\end{equation}
Hence two modes $i,j$ contribute to the commutator only if they are
non-degenerate ($\lambda_{i}\neq\lambda_{j}$) \emph{and} Euclidean
non-orthogonal ($G_{ij}\neq 0$). In the mode representation, \eqref{eq:appN-master} therefore decouples entry by entry,
\begin{equation}
    \left[2\xi^{2}(\lambda_{i}+\lambda_{j})
    + m(\lambda_{i}-\lambda_{j})^{2}\right]\hat{C}_{XV,ij}
    = 2k_{B}\tilde{T}\,\xi\,(\lambda_{i}-\lambda_{j})\,G_{ij},
\end{equation}
and back substitution in ~\eqref{eq:appN-Cv}-\eqref{eq:appN-Cx}
gives every block in closed form. We introduce the pairwise coefficient
\begin{equation}
    \mathcal{N}_{ij} \equiv 2\xi^{2}\big(\lambda_{i}+\lambda_{j}\big)
    + m\big(\lambda_{i}-\lambda_{j}\big)^{2} \;>\; 0,
    \label{eq:appN-Nij}
\end{equation}
where the inequality follows from positivity of the eigenvalues. We get an expression for the covariance matrix in terms of the eigenvalues $\{\lambda_i\}$ and the transformed diffusion matrix $G$,
\begin{subequations} \label{eq:appN-solution}
    \begin{align}
     \hat{C}_{XV,ij} &= \frac{2k_{B}\tilde{T}\,\xi\,(\lambda_{i}-\lambda_{j})}{\mathcal{N}_{ij}}\,G_{ij}, \\
     \hat C_{VV,ij} &= \frac{k_B\tilde{T}}{m}G_{ij} - \frac{\lambda_i-\lambda_j}{2\xi}\hat C_{XV,ij}
 = k_B\tilde{T}\,G_{ij}\,\frac{\mathcal N_{ij}-m(\lambda_i-\lambda_j)^2}{m\,\mathcal N_{ij}}
 = \frac{2k_B\tilde{T}\xi^2(\lambda_i+\lambda_j)}{m\,\mathcal N_{ij}}\,G_{ij}, \\
\hat{C}_{XX,ij} &= \frac{2m \hat{C}_{VV,ij}}{\lambda_i + \lambda_j} = \frac{4k_{B}\tilde{T}\,\xi^{2}}{\mathcal{N}_{ij}}\,G_{ij}.
    \end{align}
\end{subequations}

The structure in this basis is transparent: each pair of modes responds independently, with circulation proportional to its
stiffness \emph{difference} ($\hat{C}_{XV,ij} = \frac{\lambda_{i}-\lambda_{j}}{2\xi}\hat{C}_{XX,ij}$) and kinetic covariance proportional to its stiffness \emph{sum} ($\hat{C}_{VV,ij} = \frac{\lambda_{i}+\lambda_{j}}{2m}\hat{C}_{XX,ij}$), the strength of each pair's response being set by its own susceptibility $2k_{B}\tilde{T}\xi/\mathcal{N}_{ij}$. In equilibrium $[\Lambda,G]=0$, so $G_{ij}\neq0$ only between degenerate modes (this follows immediately from Eq.~\ref{eq:appN-source}). The solution then reduces to its equilibrium value, $\tilde{C}_{XV} = 0$, $\tilde{C}_{VV} = (k_{B}\tilde{T}/m)\mathbb{I}$, and $\tilde{C}_{XX} = k_{B}\tilde{T}(\mathcal{A}\tilde{\mathcal{U}})^{-1}$, as expected. 

The entropy production rate follows from $\langle\dot{\tilde{W}}\rangle = \langle\tilde{\mathbf{F}}\cdot\mathbf{V}\rangle = \langle -\tilde{\Phi}\mathbf{X}\cdot\mathbf{V}\rangle = -\mathrm{Tr}(\tilde{\Phi} \tilde{C}_{XV})$,
\begin{equation}
    \langle\dot{\tilde{S}}\rangle = \frac{\langle \dot{\tilde{W}}\rangle}{\tilde{T}}
    = -\frac{1}{2\tilde{T}}\,
    \mathrm{Tr}\left([\mathcal{A},\tilde{\mathcal{U}}]\,\tilde{C}_{XV}\right)
    = -2k_{B}\xi\sum_{i<j}
    \frac{(\lambda_{i}-\lambda_{j})^{2}\,G_{ij}\,(G^{-1})_{ij}}
    {\mathcal{N}_{ij}}.
    \label{eq:appN-EPR}
\end{equation}
Again, we emphasize that if the force matrix $\tilde{\Phi}$ is symmetric, both $[\mathcal{A},\tilde{\mathcal{U}}]$ and $\tilde{C}_{XV}$ vanish, and consequently so does the entropy production rate.

\subsubsection*{Two dimensions: the Brownian gyrator revisited}
For $N = 2$ there is a single mode pair, $(1,2)$, and the solution
simplifies drastically. The antisymmetric matrices form a one-dimensional
space, so $\tilde{C}_{XV}$, the commutator, and the mode coefficient
$\hat{C}_{XV,12}$ all reduce to single numbers, and
\eqref{eq:appN-solution}a transforms as
\begin{equation}
    {C}_{xv} = \frac{2k_{B}\tilde{T}\,\xi}{\mathcal{N}_{12}} [\mathcal{A},\tilde{\mathcal{U}}],
    \label{eq:appN-2d}
\end{equation}
where the normalization is obtained immediately from $\lambda_{1}+\lambda_{2} = \mathrm{Tr}(\mathcal{A}\tilde{\mathcal{U}})$ and $(\lambda_{1}-\lambda_{2})^{2} = (\mathrm{Tr}\,\mathcal{A}\tilde{\mathcal{U}})^{2} - 4\det(\mathcal{A}\tilde{\mathcal{U}})$, so that
\begin{equation}
    \mathcal{N}_{12} = 2\xi^{2}\,\mathrm{Tr}(\mathcal{A}\tilde{\mathcal{U}})
    + m\left[\left(\mathrm{Tr}\,\mathcal{A}\tilde{\mathcal{U}}\right)^{2}
    - 4\det(\mathcal{A}\tilde{\mathcal{U}})\right],
\end{equation}
For the Brownian gyrator the mode stiffnesses are $\lambda_{\pm} = k \pm u$ (the
normal modes of the dual gyrator potential along its principal diagonals),
giving $\mathcal{N}_{12} = 4(k\xi^{2}+mu^{2})$.

\bibliography{main}